\documentclass[12pt, draftclsnofoot, onecolumn]{IEEEtran}
\pdfoutput=1
\usepackage{amsmath,epsfig,amssymb,verbatim,amsopn,subfigure,color}
\usepackage{cite,xspace}
\usepackage{array,algorithm,algorithmic}
\usepackage{bbm}
\usepackage{array}
\usepackage{multirow}
\usepackage{multicol}
\usepackage{rotating}
\usepackage{dsfont,float}
\usepackage{stfloats}
\usepackage{fancyhdr}
\usepackage{graphicx}
\usepackage{enumerate,makecell}
\usepackage{xcolor}
\usepackage{datetime}
\usepackage{titlesec}
\allowdisplaybreaks

\titlespacing{\section}{0pt}{5pt}{-\parskip}
\normalsize

\ifCLASSOPTIONpeerreview
\renewcommand{\baselinestretch}{1.9}
\fi

\ifCLASSOPTIONonecolumn
    
\else
    
\fi

\makeatletter
\let\l@ENGLISH\l@english
\makeatother

\makeatletter
\renewcommand*{\@opargbegintheorem}[3]{\trivlist
  \item[\hskip \labelsep{\itshape #1\ #2}] {\itshape (#3):} {\normalfont}}
\makeatother
\usepackage{relsize}
\newcommand{\dGUEhori}  {r_{\textrm{T}}}
\newcommand{\dAUEhori} {r_{\textrm{A}}}

\newcommand{\AUEazi} {\phi_{\textrm{A}}}
\newcommand{\hAUE} {h_{\textrm{A}}}
\newcommand{\hBS} {h_{\textrm{BS}}}
\newcommand{\dGUEeucl} {d_{\textrm{T}}}
\newcommand{\dAUEeucl} {d_{\textrm{A}}}
\newcommand{\fadingGUE} {H_{\textrm{T}}}
\newcommand{\fadingAUE} {H_{\textrm{A}}}
\newcommand{\GUEantenna} {G_{\textrm{T}}}
\newcommand{\AUEantenna} {G_{\textrm{A}}}
\newcommand{\BSsensitivity} {\rho_{\textrm{min}}}
\newcommand{\GUEavgRpow} {\rho_{\textrm{T}}}
\newcommand{\GUETranPow} {P_{\textrm{T}}}
\newcommand{\AUETranPow} {P_{\textrm{A}}}
\newcommand{\GUERecPow} {\psi_{\textrm{T}}}
\newcommand{\AUERecPow} {\psi_{\textrm{A}}}
\newcommand{\AUEspeed} {v_{\textrm{A}}}
\newcommand{\alphaGUE} {\alpha_{\textrm{T}}}
\newcommand{\alphaLOS} {\alpha_{\textrm{L}}}
\newcommand{\alphaNLOS} {\alpha_{\textrm{N}}}
\newcommand{\rounds} {m}
\newcommand{\cellrad} {R}
\newcommand{\TPeriod} {T_{\textrm{A}}}

\newcommand{\mLOS} {m_{\textrm{L}}}
\newcommand{\mNLOS} {m_{\textrm{N}}}
\newcommand{\etaLOS} {\eta_{\textrm{L}}}
\newcommand{\etaNLOS} {\eta_{\textrm{N}}}
\newcommand{\SINRthresholdAUE} {\theta_{\textrm{A}}}
\newcommand{\SINRthresholdGUE} {\theta_{\textrm{T}}}

\newcommand{\noise} {\sigma^2}
\newcommand{\AWGN} {\mathfrak{n}}
\newcommand{\PLOS}{\mathbb{P}_{\textrm{LoS}}}
\newcommand{\muG} {\mu}
\newcommand{\BetaL} {\beta_{\textrm{L}}}
\newcommand{\BetaN} {\beta_{\textrm{N}}}
\newcommand{\NPoints}{N}
\newcommand{\ndAUEhori} {{r_{\textrm{A},\textit{n}}}}

\newcommand{\pathlossAUE} {\zeta_{\textrm{A}}}
\newcommand{\AUEsignal} {\Psi_{\textrm{A}}}
\newcommand{\GUEsignal} {\Psi_{\textrm{T}}}
\newcommand{\BSsignal} {\Psi_{\textrm{BS}}}
\newcommand{\bandwidth} {B}
\newcommand{\xGUE} {x_{\textrm{T}}}
\newcommand{\yGUE} {y_{\textrm{T}}}
\newcommand{\nxAUE} {x_{\textrm{A,\textit{n}}}}
\newcommand{\nyAUE} {y_{\textrm{A,\textit{n}}}}

\newtheorem{lemma}{Lemma}
\newtheorem{remark}{Remark}

\newtheorem{proposition}{Proposition}

\newtheorem{definition}{Definition}
\graphicspath{{figs/},{Figures/}}

\newcommand{\AuthorOne}{Nilupuli~Senadhira, {\em{Student Member, IEEE}}}
\newcommand{\AuthorTwo}{Salman~Durrani, {\em{Senior Member, IEEE}}}
\newcommand{\AuthorThree}{Xiangyun~Zhou, {\em{Senior Member, IEEE}}}
\newcommand{\AuthorFour}{Nan~Yang, {\em{Senior Member, IEEE}}}
\newcommand{\AuthorFive}{Ming~Ding, {\em{Senior Member, IEEE}}}
\newcommand{\ThankOne}{N. Senadhira, S. Durrani, X. Zhou, and N. Yang are with the Research School of Electrical, Energy and Materials Engineering, College of Engineering and Computer Science, The Australian National University, Canberra, Australia (Emails: \{nilupuli.senadhira, salman.durrani, xiangyun.zhou, nan.yang\}@anu.edu.au). M. Ding is with Data61, CSIRO, Australia (Email: ming.ding@data61.csiro.au).

Part of this work has been submitted in IEEE ICC 2020 for possible conference publication \cite{senadhira2019impact}.}

\begin{document}
\title{Uplink NOMA for Cellular-Connected UAV: Impact of UAV Trajectories and Altitude}
\author{\IEEEauthorblockN{\AuthorOne,~\AuthorTwo,~\AuthorThree,~\AuthorFour, and \AuthorFive\thanks{\ThankOne}}}
\maketitle
\setcounter{page}{1}

\begin{abstract}
This paper considers an emerging cellular-connected unmanned aerial vehicle (UAV) architecture for surveillance or monitoring applications. We consider a scenario where a cellular-connected aerial user equipment (AUE) periodically transmits in uplink, with a given data rate requirement, while moving along a given trajectory. For efficient spectrum usage, we enable the concurrent uplink transmission of the AUE and a terrestrial user equipment (TUE) by employing power-domain aerial-terrestrial non-orthogonal multiple access (NOMA), while accounting for the AUE's known trajectory. To characterize the system performance, we develop an analytical framework to compute the rate coverage probability, i.e., the probability that the achievable data rate of both the AUE and TUE exceeds the respective target rates. We use our analytical results to numerically determine the minimum height that the AUE needs to fly, at each transmission point along a given trajectory, in order to satisfy a certain quality of service (QoS) constraint for various AUE target data rates and different built-up environments. Specifically, the results show that the minimum height of the AUE depends on its distance from the BS as the AUE moves along the given trajectory. Our results highlight the importance of modeling AUE trajectory in cellular-connected UAV systems.
\end{abstract}


\ifCLASSOPTIONpeerreview
   \newpage
\fi

\newpage
\section{Introduction}
Due to their low cost, ease of deployment and mobility, unmanned aerial vehicles (UAVs) or drones are envisaged to find ever increasing use cases in civilian and commercial applications such as surveillance and monitoring, remote sensing, goods delivery, aerial photography and transporting human organs across a city \cite{geraci2019preparing}. Developing cellular support for UAVs has, therefore, been a hot topic of recent research \cite{zeng2019accessing,mozaffari2019tutorial,lin2018sky,fotouhi2019survey,zeng2018cellular}. Depending on whether UAVs are used as base stations to assist wireless communication of ground users, or as aerial users, there are two main paradigms for incorporating UAVs into cellular networks \cite{zeng2019accessing}: (i) UAV-assisted wireless communications and (ii) cellular-connected UAVs. UAV-assisted wireless communications, where UAVs act as aerial base stations (BSs), access points, relays, data aggregators, are regarded as a promising futuristic paradigm. However, before leaping  forward with the deployment of UAV BSs, cellular-connected UAVs as user equipments (UEs) are being considered by standardization bodies as a priority for now \cite{zeng2018cellular}. For instance, the 3rd Generation Partnership Project (3GPP) is expected to consider cellular-connected UAVs in Release-17, which is expected to be finalized by September 2021 \cite{geraci2019preparing}. Hence, how to seamlessly and efficiently integrate aerial user equipments (AUEs) into existing cellular networks with terrestrial UEs becomes an important problem.

Trials by a number of leading industry vendors have demonstrated the feasibility of using existing cellular networks to support UAVs in the low-altitude airspace \cite{nokia2016fcell,huawei2017digital}. In particular, it has been shown that terrestrial BSs with downtilted antennas provide adaquate coverage via antenna side lobes for UAVs flying below 120 m \cite{fotouhi2019survey}. While the antenna side lobes have reduced gain, this is offset by favorable line-of-sight (LoS) links. However, at higher altitudes, the gain from LoS links diminishes and the coverage becomes insufficient. In \cite{amer2018caching}, the use of co-ordinated multi-point (CoMP) transmission was investigated for serving aerial users, which was shown to increase the coverage up to 200 m altitude. The results in \cite{nokia2016fcell,huawei2017digital,amer2018caching,lin2018sky} assumed dedicated resources allocated to an aerial UE. However, this may not be efficient in future wireless networks as the resources dedicated to aerial users may be underutilized \cite{lin2018sky}.

Recently, non-orthogonal multiple access (NOMA) has been proposed as a promising technology to address the resource scarcity issue in wireless communications \cite{dai2018survey}. The basic idea of NOMA is to serve multiple users in the same resource block, leading to more efficient resource utilization compared to conventional orthogonal multiple access \cite{liaqat2018power,su2018power}. The use of NOMA in UAV-assisted wireless communication has been widely explored \cite{liu2019uav,
rupasinghe2018non,duan2019resource
,zhao2019joint,sun2019cyclical}. These studies focused on UAVs serving terrestrial users using NOMA in both microwave \cite{liu2019uav} and mmwave \cite{rupasinghe2018non} bands, resource allocation \cite{duan2019resource} or trajectory optimization problems \cite{liu2019uav,zhao2019joint,sun2019cyclical}. \textit{On the other hand, only a few studies have investigated the use of NOMA in cellular-connected UAV networks \cite{liu2018exploiting,mei2018uplink,rahmati2019energy}}. The energy efficiency of a NOMA scheme in a downlink mmwave cellular-connected UAV network was investigated and optimized in \cite{rahmati2019energy}. A beamforming strategy exploiting NOMA technique was proposed in \cite{liu2018exploiting} to allow an UAV to send its data to a selected subset of terrestrial BSs, which decode the UAV signal and then cancel it before decoding the messages of their served terrestrial users. By assuming the co-operative interference cancellation among the terrestrial BSs via backhaul links, a co-operative NOMA scheme for cellular-connected UAVs was proposed in \cite{mei2018uplink}. However, the coexistence of terrestrial and aerial UEs was not considered in \cite{rahmati2019energy}. In addition, the UAV mobility was not considered in \cite{rahmati2019energy,liu2018exploiting,mei2018uplink}.

Mobility is an intrinsic characteristic of UAVs, which must be properly accounted for in the modeling and design of UAV systems. In UAV-assisted wireless networks, different approaches have been taken to model the mobility of aerial BSs. Studies adopting optimization techniques focused on optimal positioning \cite{lyu2017placement,alzenad2018placement} or trajectory planning \cite{choudhury2019trajectory,huang2018cognitive,yang2018energy} problems and account for UAV mobility. Studies adopting stochastic geometry techniques typically assumed stationary UAVs and modeled them using Poisson or Binomial Point processes \cite{zhang2017spectrum,chetlur2017downlink,
al2014optimal,zhang2015antenna,lopez2018on}. Recently, some progress has been made in mobility modeling of UAVs. Two trajectory processes for aerial BSs that provide the same uniform coverage behavior as the uniform Binomial Point Process were proposed in \cite{enayati2019moving}. Inspired from the 3GPP UAV mobility model, which models UAV mobility in a straight line at a fixed height with a fixed speed, Random walk and Random waypoint models have been used to model the UAV movement in horizontal and/ or vertical directions \cite{kim2019coverage,banagar2019inspired}. However, the system performance in \cite{kim2019coverage,banagar2019inspired} is characterized from a time-average perspective, rather than the location of the UAV at a particular point along its trajectory. For cellular-connected UAVs, the prior relevant studies in \cite{rahmati2019energy,liu2018exploiting,mei2018uplink} have considered stationary aerial users only. \textit{To the best knowledge of the authors, the use of NOMA to efficiently serve both aerial and terrestrial users, while accounting for mobility of cellular-connected UAVs, has not been investigated in the literature to date.}

\subsection{Paper Contributions}
In this work, we consider a cellular-connected aerial UE (AUE) employed for surveillance or monitoring a cellular region. The AUE periodically transmits its data to the BS with a target data rate requirement. For efficient spectrum usage, we enable the concurrent uplink transmissions by the AUE and a terrestrial UE (TUE) by employing power-domain aerial-terrestrial uplink NOMA. The main contributions of this work are as follows:
\begin{itemize}
  \item We consider a successive interference cancellation (SIC) decoding strategy where the BS decodes the AUE first by treating the TUE as interference. If the AUE cannot be decoded then the BS tries to decode the TUE first and then the AUE. This decoding order leverages the fact that generally AUE link is stronger than the TUE one due to the favorable line-of-sight propagation. However, if the AUE is flying close to the cell edge and the TUE is located close to the BS, then the TUE link may be stronger than the AUE link.
  \item Using stochastic geometry, we develop a general analytical framework to compute the rate coverage probability, i.e., the probability that the achievable data rates of both the AUE and TUE exceeds the respective threshold target rates. The proposed framework explicitly incorporates the given trajectory of a cellular-connected AUE. Using the proposed framework, we determine the minimum height of the AUE for each transmission point along its given trajectory in order to meet a certain quality of service (QoS) in terms of the coverage probability for different built-up environments.
  \item Our results show that for a spiral trajectory, the minimum height increases as the AUE moves from the center to the cell edge. For low to moderate AUE target data rates, the target QoS is satisfied along the entire trajectory for suburban, urban and dense urban environments, and along the initial part of the trajectory for the urban high-rise environment. For the trajectory model adapted from the 3GPP recommendations, the minimum height of the AUE depends on its distance from the BS. The results highlight the importance of modeling the mobility related aspects in the design of NOMA-assisted cellular connected AUEs.
\end{itemize}

\subsection{Notation and Paper Organization}
The following notation is adopted in this paper. $\mathbb{P}(\cdot)$ and $\mathbb{E}[\cdot]$, denote the probability and expectation operator, respectively. $f_{X}(x)$ and $F_{X}(x)$ denote the probability density function (PDF) and cumulative distributive function (CDF) of a random variable $X$, respectively. $\lfloor\cdot\rfloor$ indicates the floor function.

The rest of the paper is organized as follows. Section \ref{sec:system_model} details the system model and the assumptions. Section \ref{sec:problem_form} describes the proposed NOMA scheme which enables the simultaneous uplink transmissions of the AUE and the paired TUE. Section \ref{sec:analytical} presents the analytical framework that is used to compute the rate coverage probabilities. Section \ref{sec:results} presents the results and highlights the impact of the system parameters on the system performance. Finally, Section \ref{sec:conclusion} concludes the paper.

\section{System Model}\label{sec:system_model}
We consider a single-cell wireless communication system with a terrestrial base station (BS), multiple stationary terrestrial user equipment (TUEs), and an aerial user equipment (AUE). The TUEs transmit in the uplink to the BS in a time division multiple access (TDMA) fashion, where each TUE in the cell is assigned a random orthogonal time slot to transmit its data to the BS. Hence, there is no intra-cell interference among the TUEs in the cell. The AUE is employed for surveillance or monitoring the region. Depending on the nature of the surveillance, the AUE has a target data rate requirement. It periodically transmits its data to the BS. For efficient spectrum usage, we assume that the AUE is not assigned a dedicated time slot for its uplink transmission. Hence, the AUE's uplink transmission interferes with the TUE transmitting at the same time. In this work, we propose to conduct uplink communication between the AUE and BS by pairing the transmissions of the AUE and a random TUE using non-orthogonal multiple access (NOMA). The TUE paired with the AUE is referred to as the \textit{active} TUE. We assume that the active TUEs in the neighbouring cells are assigned orthogonal resource blocks (RBs) to mitigate terrestrial inter-cell interference \cite{mei2018uplink}. Moreover, the AUE transmits at the same RB as the active TUE. Thus, in this work, we focus on the scenario with a single BS and a single AUE in a cell, without the presence of inter-cell interference.

\subsection{Spatial Model}
The single-cell is modeled as a disk region $\mathcal{S}$ with a radius $\cellrad$. Without loss of generality, we adopt the three-dimensional (3D) cartesian coordinate system in this work. The BS is located at the center of the cell at a fixed height $\hBS$, i.e., at coordinates $(0,0,\hBS)$. In this work, we focus on the AUE pairing with a random TUE. Hence, an assumption on the number of TUEs is not required. We assume that the active TUE is randomly located in the disk region at coordinates $(\xGUE,\yGUE,0)$ where $\dGUEhori=\sqrt{{\xGUE}^2+{\yGUE}^2}$ is the horizontal distance between the BS and the active TUE.

\begin{figure}[t]
  \centering
  \includegraphics[width=0.5\linewidth]{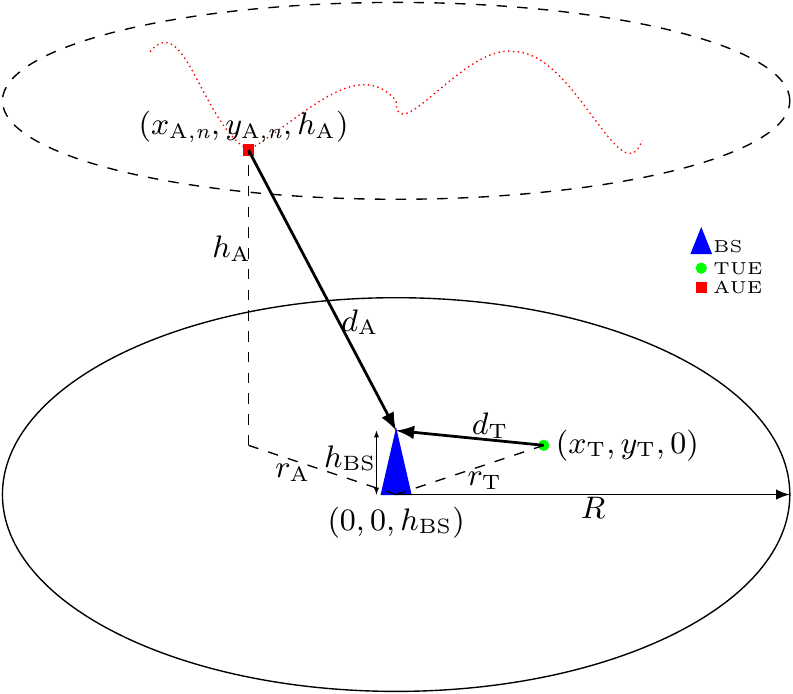}
  \caption{Illustration of the three-dimensional system model. Blue triangle, red square, and green circle represent the BS, AUE, and TUE, respectively. AUE's arbitrary trajectory is denoted by the red dotted line.}\label{fig:sm1}
\end{figure}

\subsection{AUE Trajectory and Mobility Model}
The AUE trajectory and mobility model is illustrated in Fig. \ref{fig:sm1}. We assume that the AUE flies along a given trajectory $\mathcal{T}$ to cover the disk region, with a constant speed $\AUEspeed$ at an altitude $\hAUE$ above the ground\footnote{Note that the AUE is capable of varying its height along the course of the trajectory. In Sections \ref{sec:results} C-D we assume that the AUE flies at a constant height along the entire trajectory and, in Sections \ref{sec:results} E-F, we consider a case where the AUE can vary its height at each trajectory point to achieve a certain quality of service.}. The AUE's uplink transmissions are periodic with time period $\TPeriod$, resulting in $\NPoints$ transmission points along the trajectory $\mathcal{T}$, i.e. , $\mathcal{T}\triangleq\{\mathcal{T}[n]\}_{n=1}^{\NPoints}$ where $n$ is the transmission point index. In $\mathcal{T}$, we define $\NPoints$ as $\NPoints=\lfloor\frac{s(\mathcal{T}[N])}{\AUEspeed\TPeriod}\rfloor$, where $s(\mathcal{T}[N])$ is the total path length of $\mathcal{T}$ and $\lfloor\cdot\rfloor$ is the floor function. $\mathcal{T}[n]=(\nxAUE,\nyAUE,\hAUE)$ denotes the cartesian coordinates of the location of the AUE at the $n{\textrm{th}}$ transmission point, where  $\ndAUEhori=\sqrt{{\nxAUE}^2+{\nyAUE}^2}$ is the horizontal distance  between the projection of the AUE on the ground and the BS.

\subsection{Channel Model}
Recently, progress has been made in the understanding of different types of channels in UAV communications \cite{khuwaja2018survey,zhou2018underlay,al2014optimal}. The different channels can include, air-to-air channel model (between UAVs in the sky), air-to-ground/ ground-to-air channel model (between a UAV and an user on the ground) and cellular-to-air/ air-to-cellular channel model (between a UAV and an elevated terrestrial BS). Note that the distinction between air-to-ground and air-to-cellular lies in the non-negligible height of the terrestrial BS \cite{khuwaja2018survey,zhou2018underlay,al2014optimal}.

In our system model we have the following two types of links: (i) air-to-cellular (A2C) channel between the AUE and the terrestrial BS and (ii) terrestrial channel between the active TUE and BS.

The terrestrial channel is modeled as a combination of a large-scale path-loss attenuation, with path-loss exponent $\alphaGUE$, and small-scale Rayleigh fading component, with fading power gain $\fadingGUE$. Due to the path-loss, the transmit signal power of TUE decays at a rate ${\dGUEeucl}^{-\alphaGUE}$, where $\dGUEeucl=\sqrt{{\dGUEhori}^{2}+{\hBS}^{2}}$ is the 3D propagation distance between the TUE and the BS.

Following the state-of-the-art in \cite{khuwaja2018survey,zhou2018underlay,al2014optimal}, the air-to-cellular (A2C) channel is modeled as a combination of a probabilistic distance and height dependent large-scale path-loss and small-scale Nakagami-$m$ fading, with fading power gain $\fadingAUE$. The path loss is determined according to whether the A2C channel is line-of-sight (LoS) or non-line-of-sight (NLoS) with probabilities of occurrence $\PLOS$ and $1-\PLOS$, respectively. In this work, we consider the International Telecommunication Union (ITU) model for determining the probability of LoS. The details are presented in Section  \ref{sec:results}. The corresponding path-loss is given as
\begin{equation}\label{eq:pathloss_AUE}
  \pathlossAUE=
  \begin{cases}
    \etaLOS{\dAUEeucl}^{-\alphaLOS}, & \text{if } $\textrm{LoS}$ \\
    \etaNLOS{\dAUEeucl}^{-\alphaNLOS}, & \text{if }$\textrm{NLoS}$,
  \end{cases}
\end{equation}
\noindent where $\dAUEeucl=\sqrt{{\dAUEhori}^2+{(\hAUE-\hBS)}^2}$ is the 3D propagation distance between the BS and the AUE, $\eta_{\nu}$, $\alpha_{\nu}$, $\nu\in\{\textrm{L},\textrm{N}\}$ are the additional attenuation factors and path-loss exponents for LoS and NLoS channels. The fading parameters for the LoS and NLoS channels are denoted by $\mLOS$ and $\mNLOS$, respectively.

\subsection{Received Signal}
We assume that the TUE and AUE are equipped with single omnidirectional antennas. In order to serve both the active TUE and AUE simultaneously, the BS is equipped with a dual antenna array \cite{zhang2015antenna}, i.e., BS can simultaneously beamform towards an active TUE and AUE, with associated antenna gains $\GUEantenna$ and $\AUEantenna$, respectively\footnote{The focus of this work is not on the complicated antenna modeling. Thus, we use the ideal antenna model for simplicity.}.

The AUE transmits with a fixed transmit power $\AUETranPow$. For the active TUE, we use the truncated channel inversion power control \cite{elsawy2014stochastic,zhou2018underlay}. We assume that the BS has a receiver sensitivity of $\BSsensitivity$. Therefore, the active TUE adjusts its transmit power such that average signal power received at the BS is equal to the cutoff threshold $\GUEavgRpow$, where $\GUEavgRpow>\BSsensitivity$. Hence,
\vspace{-2mm}
\begin{equation}\label{eq:GUEPower}
  \GUETranPow=\GUEavgRpow \dGUEeucl^{\alphaGUE}.
\vspace{-2mm}
\end{equation}

Based on the aforementioned system model, the received signal at the BS due to an AUE located at coordinates $(\nxAUE,\nyAUE,\hAUE)$ and an active TUE located at coordinates $(\xGUE,\yGUE,0)$ is
\begin{align}\label{eq:RecSignal}
  \BSsignal=\sqrt{\GUETranPow{\dGUEeucl}^{-\alphaGUE}\fadingGUE\GUEantenna}\GUEsignal
  +\sqrt{\AUETranPow{\pathlossAUE}\fadingAUE\AUEantenna}\AUEsignal+\AWGN,
\end{align}
\noindent where $\AWGN$ is the additive white Gaussian noise with variance $\noise$, and $\Psi_{\varrho}$, $\varrho\in\{\textrm{T},\textrm{A}\}$ denotes the signal transmitted by active TUE and AUE, respectively.

\section{Proposed NOMA scheme}\label{sec:problem_form}
The simultaneous uplink transmissions of the AUE and the active TUE are facilitated by using power-domain aerial-terrestrial uplink NOMA with successive interference cancellation (SIC) at the BS. Power-domain NOMA utilizes the power domain for user multiplexing, allowing the active TUE and AUE to share a single time-frequency resource block \cite{liaqat2018power,su2018power}. SIC allows the signals of each user to be decoded successively at the receiver end.

\textit{In the uplink terrestrial NOMA}, the user with the strongest channel link quality is decoded first while treating the users with weaker link quality as interference. Then, the decoded signal is subtracted from the superimposed signal before decoding the users with weaker link quality. This process is continued until the user with the weakest link quality is decoded \cite{liaqat2018power}. Due to analytical tractability, most studies in terrestrial uplink NOMA evaluate the channel link quality based on the average received power of each user, e.g., \cite{tabassum2017modeling}. The channel link quality ranking assumes that the user closest to the BS has the strongest channel link quality and vice versa. Thus, the decoding order in this case is fixed and distance dependent. However, the fixed decoding order ignores the possibility where the closer user experiences severe small-scale fading and farther user experiences weaker small-scale fading. This possibility is accounted for in dynamic decoding order in terrestrial NOMA \cite{salehi2018accuracy,wang2018outage}, where the channel link quality is ranked based on the instantaneous received power of each user, which considers both large-scale path loss and small-scale fading of the terrestrial users.

In this work we consider a SIC decoding strategy with adaptive decoding order. Due to the strong LoS environment of the aerial link, we assume that the received signal power corresponding to the AUE is stronger than that of the active TUE for most of the time. Based on this assumption, we assume that the AUE is decoded first at the BS\footnote{We can also consider another decoding order case where the signal transmitted by the TUE is decoded first. The working principle of this case would be similar to the case where the signal transmitted by the AUE is decoded first. However, due to the nature of the proposed NOMA scheme, the decoding order has no effect on the system performance in terms of the metrics defined in Section \ref{sec:analytical}. Thus, in this work, we only focus on the case where the signal transmitted by the AUE is decoded first.}. However, if the AUE cannot be decoded, then the BS tries to decode the TUE, and then the AUE. This additional decoding step accounts for the case, where the received power corresponding to the active TUE is greater than that of the AUE when the AUE is flying closer to the cell edge and the active TUE is located closer to the BS. A detailed description of the decoding events and the aforementioned extra decoding step of the proposed aerial-terrestrial NOMA scheme is presented below.

The tree  diagram of the decoding events in the proposed NOMA scheme is illustrated in Fig. \ref{fig:dec_order_AUE} and explained as follows. The received signal at the BS is comprised of the superimposed $\AUEsignal$ and $\GUEsignal$ signals, where $\AUEsignal$ and $\GUEsignal$ denote the signals transmitted by the AUE and TUE, respectively. With SIC at the BS, $\AUEsignal$  is decoded first by treating $\GUEsignal$ as interference. If $\AUEsignal$ is decoded successfully, $\GUEsignal$ is decoded using SIC. Otherwise, BS tries to decode $\GUEsignal$ while treating $\AUEsignal$ as interference. In this work, we assume that, if $\AUEsignal$ is not decoded successfully, then error propagation occurs, i.e., BS treats the AUE's entire signal as interference when decoding $\GUEsignal$. If $\AUEsignal$ is is decoded successfully, the error propagation factor is 0, i.e,. AUE's entire signal is subtracted from the superimposed signal \cite{sun2016non}. If $\GUEsignal$ is decoded successfully at this stage, BS tries to decode the previously unsuccessful $\AUEsignal$ using SIC.

Each branch of the probability tree in Fig. \ref{fig:dec_order_AUE} corresponds to a joint decoding event where either/ both/ none of the signals are decoded. Events corresponding to each branch are defined as follows.
\begin{itemize}
  \item ${E_1}:$ Event that $\AUEsignal$ (i.e., AUE) is decoded in the first step and $\GUEsignal$ (i.e., TUE) is decoded in the second step.
  \item $E_2:$ Event that $\AUEsignal$ is decoded in the first step and $\GUEsignal$ is not decoded in the second step.
  \item $E_3:$ Event that $\AUEsignal$ is not decoded in the first step, $\GUEsignal$ is decoded in the second step, and $\AUEsignal$ is decoded in the third step.
  \item $E_4:$ Event that $\AUEsignal$ is not decoded in the first step, $\GUEsignal$ is decoded in the second step, and $\AUEsignal$ is not decoded in the third step.
  \item $E_5:$ Event that $\AUEsignal$ is not decoded in the first step and $\GUEsignal$ is not decoded in the second step.
\end{itemize}

\begin{remark}
  Due to the extra decoding step (corresponding to events $E_3$ and $E_4$), the tree diagram in this work is different from the tree diagrams in prior studies in terrestrial NOMA \cite{xia2018outage,tabassum2017modeling}. In addition, we take the dependency of individual decoding steps into account rather than making the assumption that the decoding steps are independent as done in \cite{tabassum2017modeling}.
\end{remark}

The probability corresponding to each event $E_{i}$ is given by $P_{i}=\mathbb{P}\left(E_{i}\right)$, where $i=1,2,3,4$. These probabilities will be derived in Section \ref{sec:analytical} to characterize the system performance under the proposed NOMA scheme.

\section{Analytical Framework}\label{sec:analytical}
In this section, we analyze the performance of the system using the rate coverage probability as the performance metric.

\begin{definition}
The rate coverage probability is the probability that the achievable data rate of a user exceeds the target data rate. It is defined as $\mathbb{P}(\bandwidth \log_2(1+\mathrm{SINR}_{\varrho})\geq\pi_{\varrho})$, where $\varrho\in\{\mathrm{T},\mathrm{A}\}$ denotes TUE and AUE, and $\bandwidth$, $\mathrm{SINR}_{\varrho}$ and $\pi_{\varrho}$ correspond to the bandwidth, signal-to-interference-plus-noise ratio and target data rate of the user, respectively.
\end{definition}

For analytical simplicity, we re-express the rate coverage probability as
\begin{align}\label{eq:SINR_ccdf}
  P_{\mathrm{cov}}=\mathbb{P}[\mathrm{SINR}_{\varrho}\geq\theta_{\varrho}],
\end{align}
\noindent which is the complementary cumulative distribution function (CCDF) of SINR, where $\theta_{\varrho}=2^{\frac{\pi_{\varrho}}{B}-1}$ is the target SINR threshold of the user. Note that we evaluate the rate coverage probability at each trajectory point of AUE, by averaging it over the location of the active TUE, and small-scale fading of AUE and TUE.

For each trajectory point, we evaluate the performance by computing three main rate coverage probabilities as defined below. The following probabilities correspond to linear combinations of probabilities of joint decoding events defined in Section \ref{sec:problem_form}.
\begin{itemize}
  \item $P_{\mathrm{Tot}}:$ Rate coverage probability of the event where both AUE and TUE are decoded successfully and is given by $P_{\mathrm{Tot}}=P_1+P_3$.
  \item $P_{\mathrm{AUE}}:$ Rate coverage probability of the event where AUE is decoded successfully and is given by $P_{\mathrm{AUE}}=P_1+P_2+P_3$.
  \item $P_{\mathrm{TUE}}:$ Rate coverage probability of the event where TUE is decoded successfully and is given by $P_{\mathrm{TUE}}=P_1+P_3+P_4$.
\end{itemize}

\begin{figure}[t]
\centering
\begin{minipage}{.45\textwidth}
  \centering
  \includegraphics[width=0.9\linewidth]{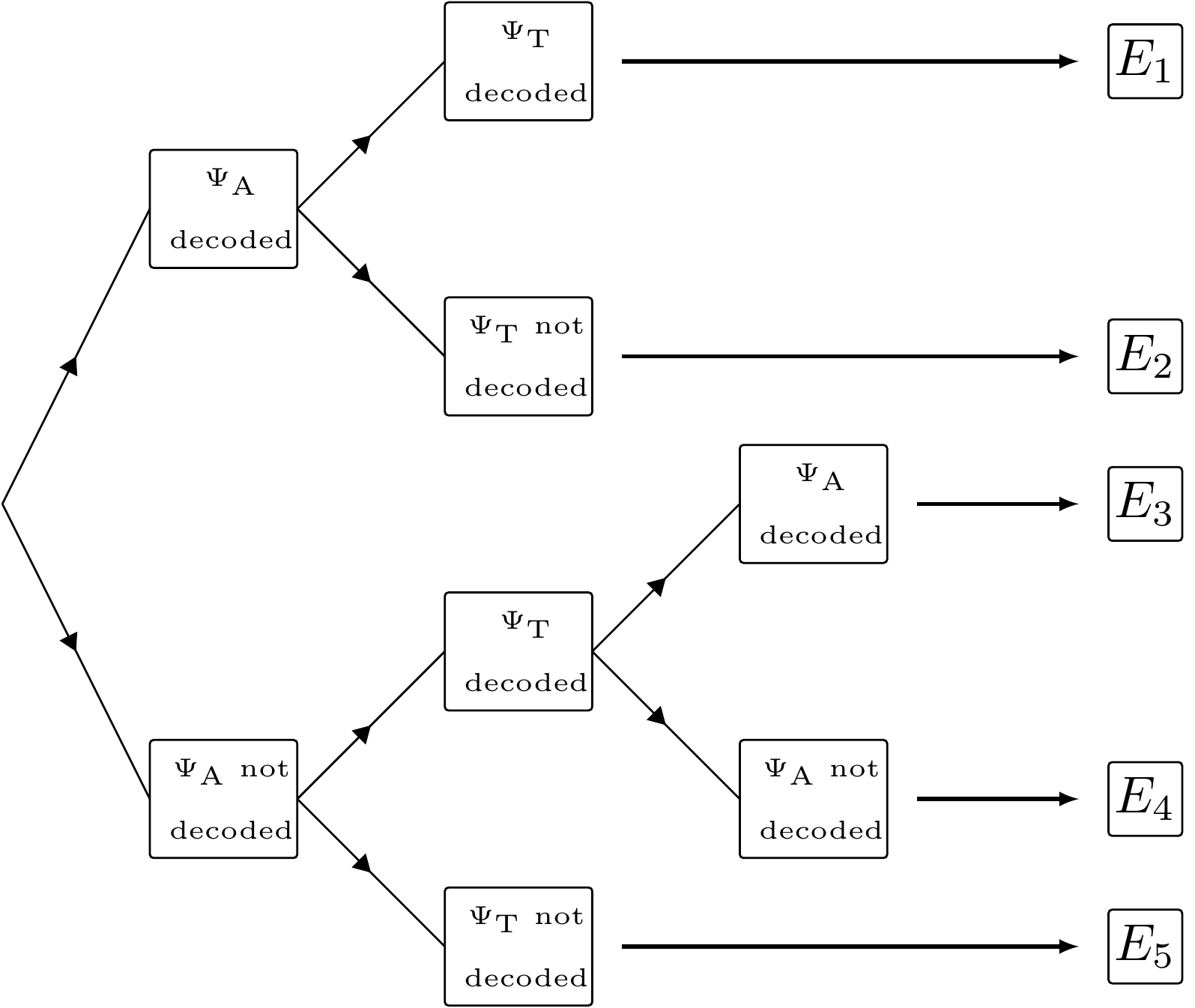}
  \caption{Tree diagram of decoding events for proposed NOMA scheme.}
  \label{fig:dec_order_AUE}
\end{minipage}%
\hspace{1 cm}
\begin{minipage}{.45\textwidth}
  \centering
  \includegraphics[scale=0.38]{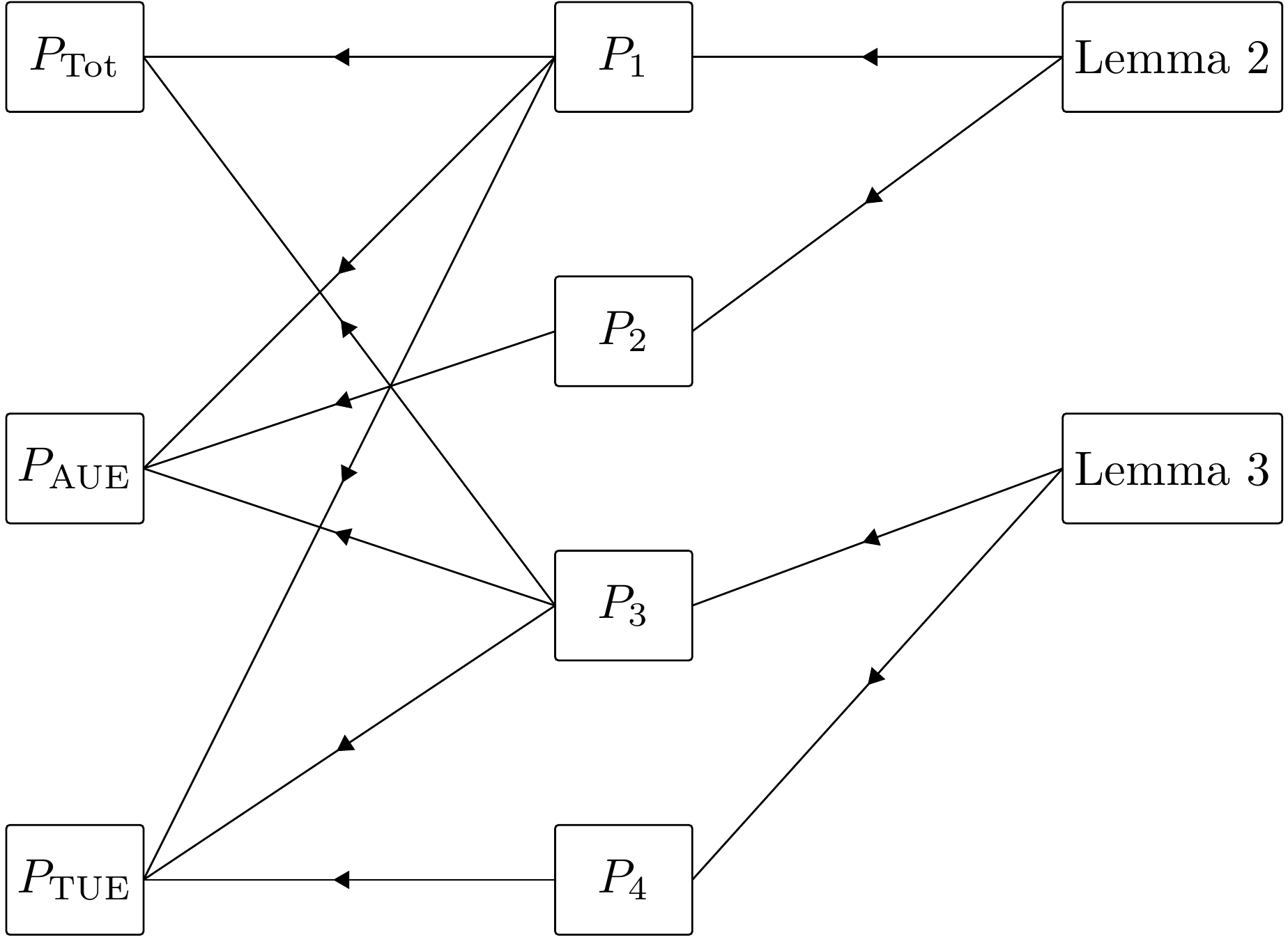}
  \caption{Summary of the analytical framework metrics.}
  \label{fig:metric_diagram}
\end{minipage}
\end{figure}

The rest of this section presents the lemmas and propositions used in the derivation of $P_{\mathrm{Tot}}$, $P_{\mathrm{AUE}}$ and $P_{\mathrm{TUE}}$. The relationship between the main results is illustrated in Fig. \ref{fig:metric_diagram}.

First we present three Lemmas, which help to derive the main results in this paper.

\begin{lemma}\label{lem:pdf_dG}
The probability density function (PDF) of the 3D propagation distance $\dGUEeucl$ between BS and TUE is
\begin{align}\label{eq:pdf_dG}
  f_{\dGUEeucl}(z)=\frac{2z}{\cellrad^2} \text{ ,       } \hBS\leq z\leq\sqrt{\cellrad^2+\hBS^2}.
\end{align}
\begin{IEEEproof}
See Appendix \ref{app:proof_lemma_dG}.
\end{IEEEproof}
\end{lemma}

\begin{lemma}\label{lem:pdf_SINRG}
The cumulative distribution function (CDF) and PDF of the received power $\GUERecPow$ corresponding to the active TUE, assuming TUE fading channel is Rayleigh fading, are
\begin{align}\label{eq:cdf:SINRG}
  F_{\GUERecPow}(x)=1-\exp\left(\frac{-x}{\GUEavgRpow\GUEantenna}\right),
\end{align}
and
\begin{align}\label{eq:pdf_SINRG}
    f_{\GUERecPow}(x)=\frac{1}{\GUEavgRpow \GUEantenna}\exp{\left(\frac{-x}{\GUEavgRpow \GUEantenna}\right)},
\end{align}
\noindent respectively, where $\GUERecPow=\GUETranPow{\dGUEeucl}^{-\alphaGUE}\fadingGUE\GUEantenna$.
\begin{IEEEproof}
See Appendix \ref{app:proof_lemma_xG_xA}.
\end{IEEEproof}
\end{lemma}

\begin{lemma}\label{lem:pdf_SINRA}
The CDF and PDF of the received power $\AUERecPow$ corresponding to the AUE are
\begin{align}\label{eq:cdf_SINRA}
  F_{\AUERecPow}(x)=1-\PLOS\sum_{i=0}^{\mLOS-1}\frac{{\left(\BetaL x\right)}^i}{i!}\exp(-\BetaL x) -
  (1-\PLOS)\sum_{j=0}^{\mNLOS-1}\frac{{\left(\BetaN x\right)}^j}{j!}\exp(-\BetaN x),
\end{align}
and
\begin{align}\label{eq:pdf_SINRA}
  f_{\AUERecPow}(x)=\PLOS\frac{\exp(-x \BetaL) {\BetaL}^{\mLOS} x^{\mLOS-1}}{\Gamma(\mLOS)} +(1- \PLOS)\frac{\exp(-x \BetaN) {\BetaN}^{\mNLOS} x^{\mNLOS-1}}{\Gamma(\mNLOS)},
\end{align}
\noindent respectively, where $\AUERecPow=\AUETranPow{\pathlossAUE}\fadingAUE\AUEantenna$, $\BetaL=\frac{\mLOS}{\AUETranPow \etaLOS \dAUEeucl^{-\alphaLOS} \AUEantenna}$ and $\BetaN=\frac{\mNLOS}{\AUETranPow \etaNLOS \dAUEeucl^{-\alphaNLOS} \AUEantenna}$.
\begin{IEEEproof}
See Appendix \ref{app:proof_lemma_xG_xA}.
\end{IEEEproof}
\end{lemma}

Next, we derive the rate coverage probabilities of the joint decoding events defined in Section \ref{sec:problem_form}.

\begin{proposition}\label{prop:P1}
The rate coverage probability that AUE is decoded in the first step and TUE is decoded in the second step is
\begin{align}\label{eq:P1} \nonumber
  P_1&=\frac{1}{\muG}\exp\left(\frac{\noise}{\muG}\right)\Biggl[
  \PLOS \sum_{i=0}^{\mLOS-1}
  \frac{{(\BetaL\SINRthresholdAUE)}^{i}}{i!}{\left(\BetaL\SINRthresholdAUE+\frac{1} {\muG}\right)}^{-i-1}\Gamma\left(1+i,(1+\SINRthresholdGUE)\left(\BetaL\SINRthresholdAUE+
  \frac{1}{\muG}\right)\noise\right)\\
  &+(1-\PLOS) \sum_{j=0}^{\mNLOS-1}
  \frac{{(\BetaN\SINRthresholdAUE)}^{j}}{i!}{\left(\BetaN\SINRthresholdAUE+\frac{1} {\muG}\right)}^{-j-1}\Gamma\left(1+j,(1+\SINRthresholdGUE)\left(\BetaN\SINRthresholdAUE+
  \frac{1}{\muG}\right)\noise\right)
  \Biggl],
\end{align}
\noindent where $\muG=\GUEavgRpow\GUEantenna$.
\begin{IEEEproof}
The proof relies on stochastic geometry and is presented in Appendix \ref{app:proof_prop_p1_p2}.
\end{IEEEproof}
\end{proposition}

\begin{proposition}\label{prop:P2}
The rate coverage probability that AUE is decoded in the first step and TUE is not
decoded in the second step is
\begin{align}\label{eq:P2} \nonumber
  P_2&=\frac{1}{\muG}\exp\left(\frac{\noise}{\muG}\right)\Biggl[
  \PLOS \sum_{i=0}^{\mLOS-1}
  \frac{{(\BetaL\SINRthresholdAUE)}^{i}}{i!}{\left(\BetaL\SINRthresholdAUE+\frac{1} {\muG}\right)}^{-i-1}\Bigg[\Gamma\left(1+i,\left(\BetaL\SINRthresholdAUE+
  \frac{1}{\muG}\right)\noise\right)\\ \nonumber
  &-\Gamma\left(1+i,(1+\SINRthresholdGUE)\left(\BetaL\SINRthresholdAUE+
  \frac{1}{\muG}\right)\noise\right)\Bigg]
  +(1-\PLOS)\sum_{j=0}^{\mNLOS-1}\frac{{(\BetaN\SINRthresholdAUE)}^{j}}{j!}
  {\left(\BetaN\SINRthresholdAUE+\frac{1}{\muG}\right)}^{-j-1}\\
  &\times\Bigg[\Gamma\left(1+j,\left(\BetaN\SINRthresholdAUE+
  \frac{1}{\muG}\right)\noise\right)-\Gamma\left(1+j,(1+\SINRthresholdGUE)
  \left(\BetaN\SINRthresholdAUE+\frac{1}{\muG}\right)\noise\right)\Bigg]
  \Biggl].
\end{align}
\begin{IEEEproof}
The proof relies on stochastic geometry and is presented in Appendix \ref{app:proof_prop_p1_p2}.
\end{IEEEproof}
\end{proposition}

\begin{proposition}\label{prop:P3}
The rate coverage probability that AUE is not decoded in the first step, TUE is decoded
in the second step, and AUE is decoded in the third step is

\begin{equation}\label{eq:P3}
  P_{3}=
  \begin{cases}
    \exp\left(\frac{-\SINRthresholdGUE\noise}{\muG}\right)\Biggl[
  \frac{\PLOS}{\Gamma(\mLOS)}\BetaL^{\mLOS}{\left(\BetaL+\frac{\SINRthresholdGUE}{\muG}
  \right)}^{-\mLOS}\Gamma\left(\mLOS,\SINRthresholdAUE\left(\BetaL+\frac{\SINRthresholdGUE}{\muG}
  \right)\noise\right)\\
  \,\,\,+\frac{1-\PLOS}{\Gamma(\mNLOS)}\BetaN^{\mNLOS}{\left(\BetaN+\frac{\SINRthresholdGUE}{\muG}
  \right)}^{-\mNLOS}\Gamma\left(\mNLOS,\SINRthresholdAUE\left(\BetaN+\frac{\SINRthresholdGUE}{\muG}
  \right)\noise\right)
  \Biggl]
 , & \mbox{if } \SINRthresholdAUE\SINRthresholdGUE\geq1 \\
   \frac{\PLOS}{\Gamma(\mLOS)}\exp\left(\frac{\noise}{\muG}\right){\BetaL}^{\mLOS}{\left(
   \BetaL+\frac{1}{\SINRthresholdAUE\muG}\right)}^{-\mLOS}\Gamma\left(
   \mLOS,\left(\BetaL\SINRthresholdAUE+\frac{1}{\muG}\right)\noise\right)\\
   \,\,\,+\frac{1-\PLOS}{\Gamma(\mNLOS)}\exp\left(\frac{\noise}{\muG}\right){\BetaN}^{\mNLOS}{\left(
   \BetaN+\frac{1}{\SINRthresholdGUE\muG}\right)}^{-\mNLOS}\Gamma\left(
   \mNLOS,\left(\BetaN\SINRthresholdAUE+\frac{1}{\muG}\right)\noise\right)
   -\frac{1}{2}\frac{\SINRthresholdAUE{\SINRthresholdGUE}^2\sigma^4
   {(1+\SINRthresholdAUE)}^2}{(1-\SINRthresholdAUE\SINRthresholdGUE)}
    , & \mbox{if } 0\leq\SINRthresholdAUE\SINRthresholdGUE<1.
  \end{cases}
\end{equation}
\begin{IEEEproof}
\begin{figure}[t]
\centering
\subfigure[$\SINRthresholdAUE \SINRthresholdGUE>1$ ]{\label{fig:P3_greater}\includegraphics[width=0.32\textwidth]{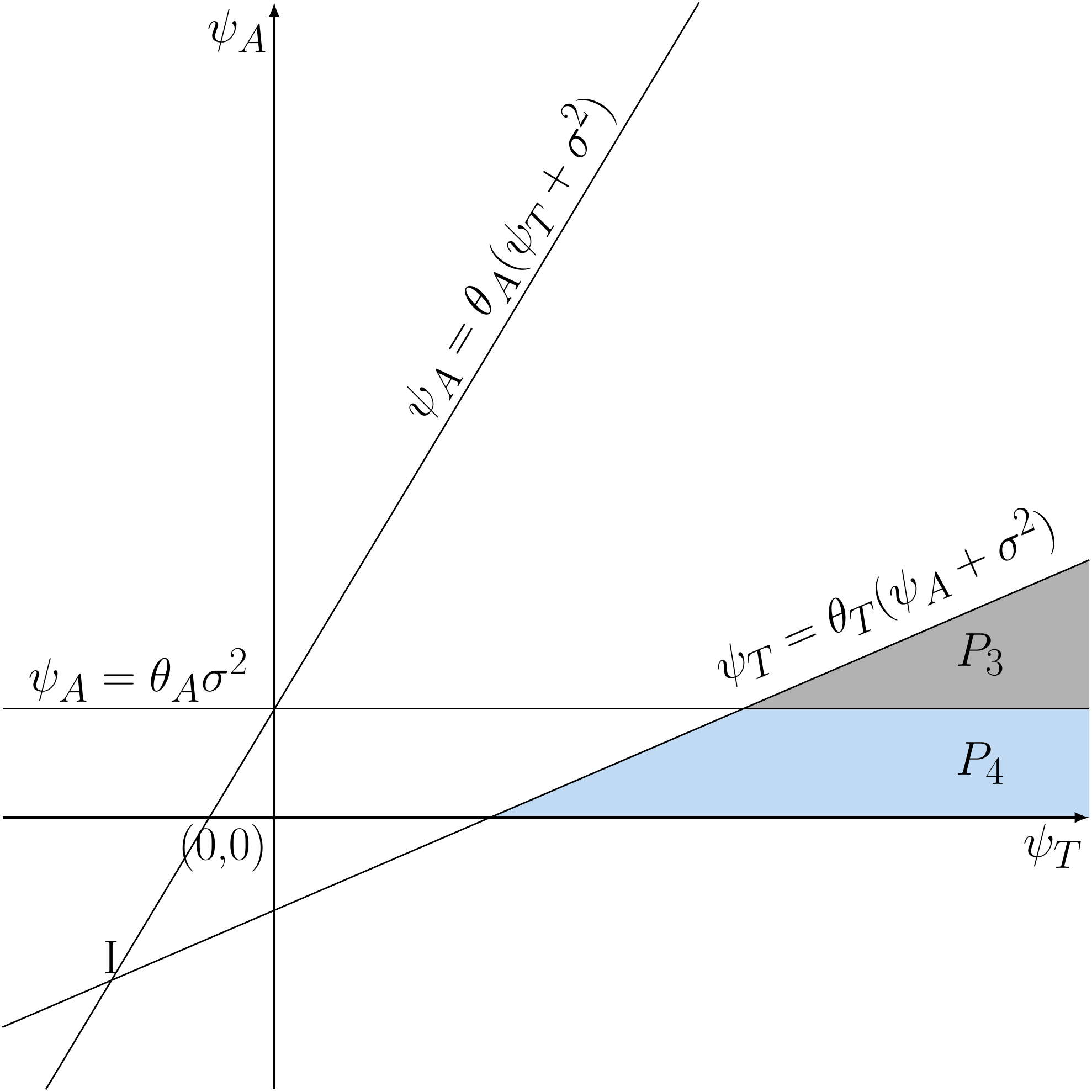}}
\subfigure[$\SINRthresholdAUE \SINRthresholdGUE=1$ ]{\label{fig:P3_equal}\includegraphics[width=0.32\textwidth]{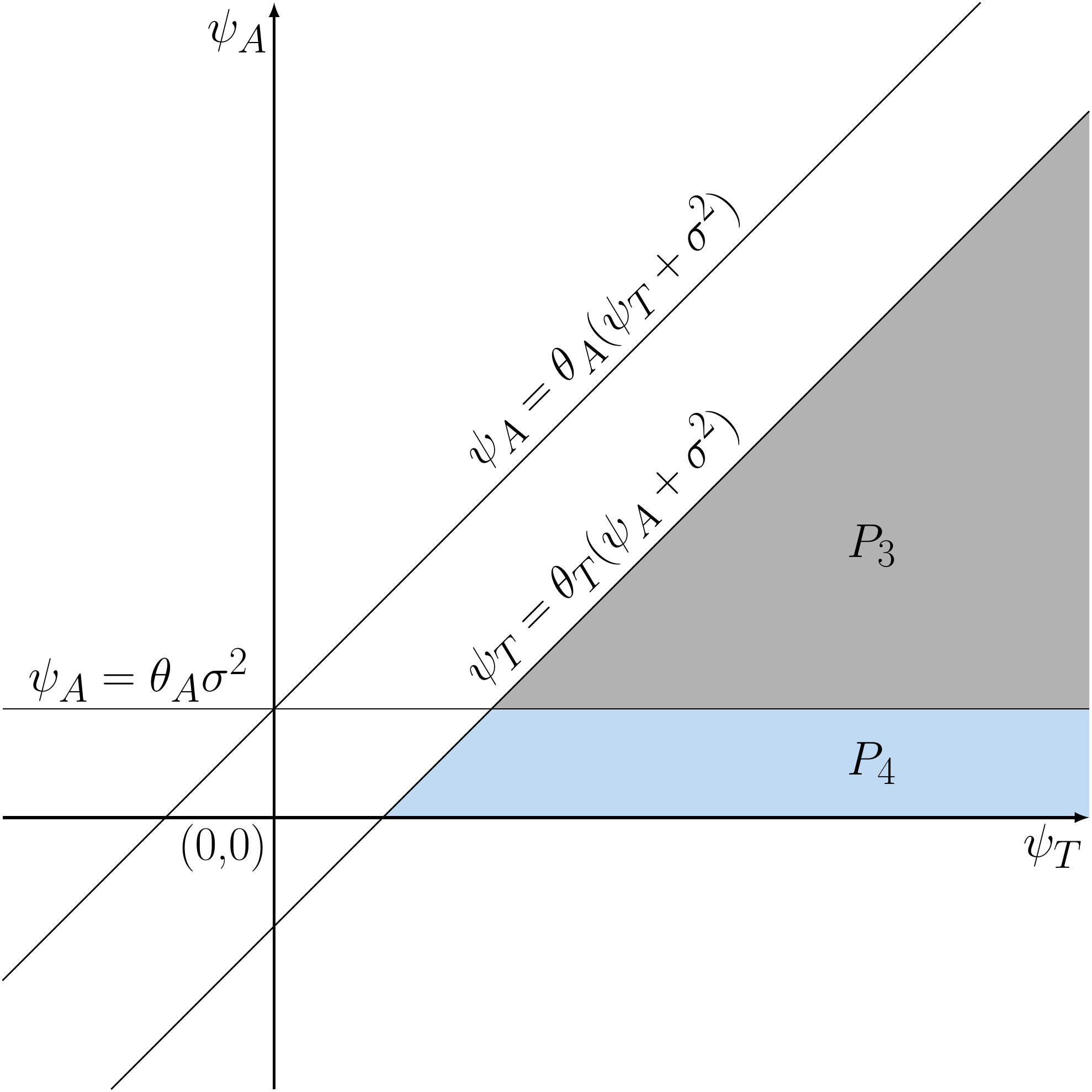}}
\subfigure[$\SINRthresholdAUE \SINRthresholdGUE<1$ ]{\label{fig:P3_less}\includegraphics[width=0.32\textwidth]{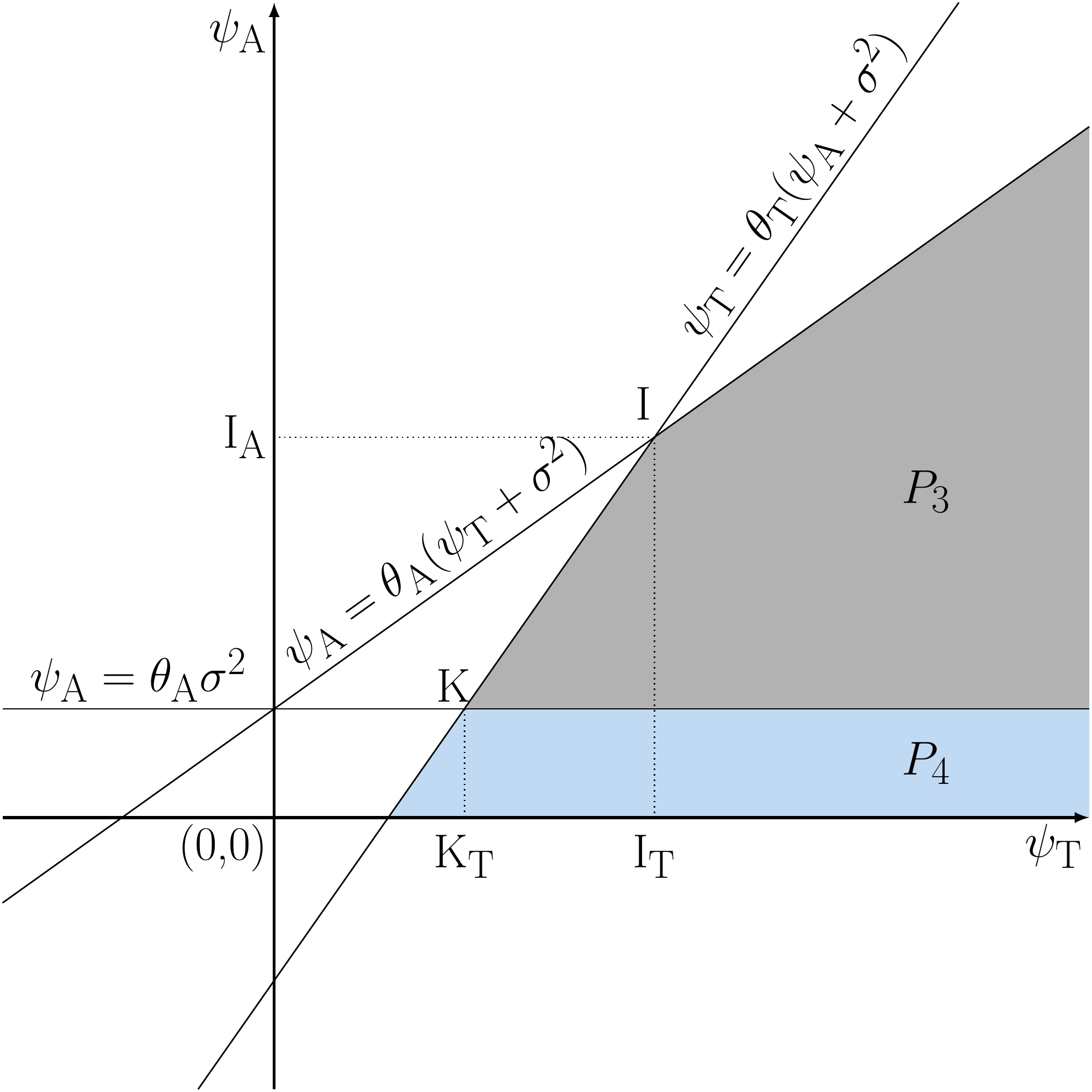}}
\caption{Integral regions of $P_3$ and $P_4$, when $\SINRthresholdAUE\SINRthresholdGUE>1$, $\SINRthresholdAUE\SINRthresholdGUE=1$, and $\SINRthresholdAUE\SINRthresholdGUE<1$, respectively, for proof of Propositions \ref{prop:P3} and \ref{prop:P4}.}\label{fig:P3_P4_plots}
\end{figure}
The rate coverage probability $P_3$ can be expressed as
\begin{subequations}\label{eq:P3_proofG}
  \begin{align}
  \nonumber
    P_3&=\mathbb{P}_{\AUERecPow,\GUERecPow}\left(\frac{\AUERecPow}{\noise}\geq
    \SINRthresholdAUE, \frac{\GUERecPow}{\AUERecPow+\noise}\geq\SINRthresholdGUE,
    \frac{\AUERecPow}{\GUERecPow+\noise}<\SINRthresholdAUE \right)\\ \label{eq:P3_proofG_3ineq}
    &=\mathbb{P}_{\AUERecPow,\GUERecPow}\left(\AUERecPow\geq\SINRthresholdAUE
    \noise, \GUERecPow\geq\SINRthresholdGUE(\AUERecPow+\noise),
    \AUERecPow<\SINRthresholdAUE(\GUERecPow+\noise)\right).
    \end{align}
    \end{subequations}
\noindent The inequalities in \eqref{eq:P3_proofG_3ineq} are plotted in Fig. \ref{fig:P3_P4_plots}. $P_3$ is derived by calculating the area covered by all three curves. The point of intersection of $\AUERecPow=\SINRthresholdAUE(\GUERecPow+\noise)$ and $\GUERecPow=\SINRthresholdGUE(\AUERecPow+\noise)$ is given by $\mathrm{I}=\left(\frac{\SINRthresholdGUE\noise(1+\SINRthresholdAUE)}
{1-\SINRthresholdAUE\SINRthresholdGUE},\frac{\SINRthresholdAUE\noise
(1+\SINRthresholdGUE)}{1-\SINRthresholdAUE\SINRthresholdGUE}\right)$. Depending on the value of $\SINRthresholdAUE\SINRthresholdGUE$, the point of intersection, $\mathrm{I}$ can be located in different quadrants or non-existent (in the case of $\SINRthresholdAUE\SINRthresholdGUE=1$). Thus, $P_3$ can have different values in these cases.

We first present the proof of $P_{3}$ when $\SINRthresholdAUE\SINRthresholdGUE\geq1$.
  \begin{subequations}\label{P3_proof_Greater}
  \begin{align}
  \label{eq:P3_proofG_2ineq}
    P_3&=\mathbb{P}_{\AUERecPow,\GUERecPow}\left(\GUERecPow\geq\SINRthresholdGUE
    (\AUERecPow+\noise), \AUERecPow\geq\SINRthresholdAUE\noise \right)\\ \nonumber
    &=\mathbb{E}_{\AUERecPow}\left[\mathbb{P}_{\GUERecPow}\left(
    \GUERecPow\geq\SINRthresholdGUE(a+\noise),a\geq\SINRthresholdAUE\noise
    \right)\right]\\ \label{eq:P3_proofG_intProb}
    &=\int_{\SINRthresholdAUE\noise}^{\infty}\mathbb{P}_{\GUERecPow}\left(
    \GUERecPow\geq\SINRthresholdGUE(a+\noise)\right) f_{\AUERecPow}(a)da \\ \nonumber
    &=\int_{\SINRthresholdAUE\noise}^{\infty}\mathbb{P}_{\fadingGUE,\dGUEeucl}
    \left(\fadingGUE\geq\frac{\GUERecPow(a+\noise)}
    {\GUETranPow{\dGUEeucl}^{-\alphaGUE}\GUEantenna}\right) f_{\AUERecPow}(a)da \\  \label{eq:P3_proofG_GUEfad}
    &=\int_{\SINRthresholdAUE\noise}^{\infty}\mathbb{E}_{\dGUEeucl}\left[\exp\left(-\frac{\SINRthresholdGUE
    (a+\noise)}{\GUETranPow{\dGUEeucl}^{-\alphaGUE}\GUEantenna}\right) \right]
    f_{\AUERecPow}(a)da \\ \nonumber
    &=\int_{\SINRthresholdAUE\noise}^{\infty}\left[\int_{\hBS}^{\sqrt{{\hBS}^2+{\cellrad}^2}}
    \exp\left(-\frac{\SINRthresholdGUE(a+\noise)}{\GUEavgRpow z^{\alphaGUE}z^{-\alphaGUE}\GUEantenna}\right)\left(\frac{2z}{\cellrad^2}\right)
    dz\right] f_{\AUERecPow}(a)da \\
    &=\int_{\SINRthresholdAUE\noise}^{\infty}\exp\left(\frac{-\SINRthresholdGUE
    (a+\noise)}{\GUEavgRpow\GUEantenna} \right)
    \left[\PLOS\frac{\exp(-a \BetaL) {\BetaL}^{\mLOS} a^{\mLOS-1}}{\Gamma(\mLOS)} +(1- \PLOS)\frac{\exp(-a \BetaN) {\BetaN}^{\mNLOS} a^{\mNLOS-1}}{\Gamma(\mNLOS)}\right] da , \label{eq:P3_proofG_int}
  \end{align}
\end{subequations}
\noindent where \eqref{eq:P3_proofG_2ineq} is the simplified expression for the area of $P_3$ based on Fig. \ref{fig:P3_greater} and \ref{fig:P3_equal}. \eqref{eq:P3_proofG_GUEfad} comes from the fact that $\fadingGUE$ follows an exponential distribution. Integration of \eqref{eq:P3_proofG_int} with respect to $a$ and substitution of $\mu=\GUEavgRpow\GUEantenna$ into \eqref{eq:P3_proofG_int} yield \eqref{eq:P3} for the case $\SINRthresholdAUE\SINRthresholdGUE\geq1$.

The proof of $P_3$ for the case $\SINRthresholdAUE\SINRthresholdGUE<1$ is presented as follows. Let $\mathrm{I}=(\mathrm{I}_{\mathrm{T}},\mathrm{I}_{\mathrm{A}})$, where ${\mathrm{I}_{\mathrm{T}}}$ denotes the x-coordinate and ${\mathrm{I}_{\mathrm{A}}}$ denotes the y-coordinate of point $\mathrm{I}$, and, $\mathrm{K}_{\mathrm{T}}=\SINRthresholdGUE\noise(\SINRthresholdAUE+1)$ is the x-coordinate of point K as illustrated in Fig. \ref{fig:P3_less}. Let $P_{3,Q}$ be the triangular area enclosed by the curves, $\GUERecPow=\SINRthresholdGUE(\AUERecPow+\noise)$, $\AUERecPow=\SINRthresholdAUE\noise$ and $\GUERecPow=\mathrm{I}_{\mathrm{T}}$. Let $P_{3,R}$ be the quadrilateral area enclosed by the curves, $\AUERecPow=\SINRthresholdAUE(\GUERecPow+\noise)$, $\AUERecPow=\SINRthresholdAUE\noise$ and $\GUERecPow=\mathrm{I}_{\mathrm{T}}$. Thus, we can write $P_{3}=P_{3,Q}+P_{3,R}$. The derivations of $P_{3,Q}$ and $P_{3,R}$ are presented below.
\begin{subequations}\label{eq:P3Q_proof}
  \begin{align}
  \label{eq:P3Q_proof_triangle}
    P_{3,Q}&=\frac{1}{2}(\mathrm{I}_{\mathrm{T}}-\mathrm{K}_{\mathrm{T}})
    (\mathrm{I}_{\mathrm{A}}-\SINRthresholdAUE\noise)
    =\frac{1}{2}\frac{\SINRthresholdAUE{\SINRthresholdGUE}^2\noise
    (1+\SINRthresholdAUE)}{(1-\SINRthresholdAUE\SINRthresholdGUE)}
    \frac{\SINRthresholdAUE\SINRthresholdGUE\noise(1+\SINRthresholdAUE)}
    {(1-\SINRthresholdAUE\SINRthresholdGUE)}\\
    &=\frac{1}{2}\frac{{\SINRthresholdAUE}^2{\SINRthresholdGUE}^3{\sigma}^4
    {(1+\SINRthresholdAUE)}^2}{{(1-\SINRthresholdAUE\SINRthresholdGUE)}^2},
  \end{align}
\end{subequations}
\noindent where \eqref{eq:P3Q_proof_triangle} is the area of a triangle.
\begin{subequations}\label{eq:P3R_proof}
  \begin{align}
  \label{eq:P3R_proof_trian_diff}
    P_{3,R}&=\mathbb{P}_{\AUERecPow,\GUERecPow}\left(\AUERecPow<\SINRthresholdAUE
    (\GUERecPow+\noise),\AUERecPow\geq\SINRthresholdAUE\noise\right)-
    \frac{1}{2}\mathrm{I}_{\mathrm{T}}(\mathrm{I}_{\mathrm{A}}-\SINRthresholdAUE\noise)\\ \nonumber
    &=\mathbb{E}_{\AUERecPow}\left[\mathbb{P}_{\GUERecPow}\left(
    \GUERecPow>\frac{a}{\SINRthresholdAUE}-\noise,a\geq\SINRthresholdAUE
    \noise\right)\right]
    -\frac{1}{2}\frac{\SINRthresholdAUE{\SINRthresholdGUE}^2
    \sigma^4{(1+\SINRthresholdAUE)}^2}{{(1-\SINRthresholdAUE\SINRthresholdGUE)}^2}\\ \nonumber
    &=\int_{\SINRthresholdAUE\noise}^{\infty}\mathbb{P}_{\GUERecPow}\left(
    \GUERecPow>\frac{a}{\SINRthresholdAUE}-\noise\right) f_{\AUERecPow}(a)da
    -\frac{1}{2}\frac{\SINRthresholdAUE{\SINRthresholdGUE}^2
    \sigma^4{(1+\SINRthresholdAUE)}^2}{{(1-\SINRthresholdAUE\SINRthresholdGUE)}^2}\\ \label{eq:P3R_proof_last}
    &=\int_{\SINRthresholdAUE\noise}^{\infty}\exp\left(-\frac{(\frac{a}
    {\SINRthresholdAUE}-\noise)}{\GUEavgRpow\GUEantenna}\right) f_{\AUERecPow}(a)da -\frac{1}{2}\frac{\SINRthresholdAUE{\SINRthresholdGUE}^2
    \sigma^4{(1+\SINRthresholdAUE)}^2}{{(1-\SINRthresholdAUE\SINRthresholdGUE)}^2},
  \end{align}
\end{subequations}
\noindent where \eqref{eq:P3R_proof_trian_diff}  is the difference between the triangular area enclosed by $\AUERecPow=\SINRthresholdAUE(\GUERecPow+\noise)$ and $\AUERecPow=\SINRthresholdAUE\noise$, and the triangular area enclosed by $\AUERecPow=\SINRthresholdAUE(\GUERecPow+\noise)$, $\AUERecPow=\SINRthresholdAUE\noise$ and $\GUERecPow=\mathrm{I}_{\mathrm{T}}$. Evaluating and simplifying \eqref{eq:P3R_proof_last}, we obtain \eqref{eq:P3}.
\end{IEEEproof}
\end{proposition}

\begin{proposition}\label{prop:P4}
The rate coverage probability that AUE is not decoded in the first step, TUE is
decoded in the second step, and AUE is not decoded in the third step is
\begin{align}\label{eq:P4} \nonumber
  P_4&=\exp\left(\frac{-\SINRthresholdGUE\noise}{\muG}\right)\Biggl[
  \frac{\PLOS}{\Gamma(\mLOS)}\BetaL^{\mLOS}{\left(\BetaL+\frac{\SINRthresholdGUE}{\muG}
  \right)}^{-\mLOS}\Bigg[\Gamma(\mLOS)-\Gamma\left(\mLOS,\SINRthresholdAUE\left(\BetaL+
  \frac{\SINRthresholdGUE}{\muG}\right)\noise\right)\Bigg]\\
  &+\frac{1-\PLOS}{\Gamma(\mNLOS)}\BetaN^{\mNLOS}{\left(\BetaN+\frac{\SINRthresholdGUE}
  {\muG}\right)}^{-\mNLOS}\Bigg[\Gamma(\mNLOS)-\Gamma\left(\mNLOS,\SINRthresholdAUE\left(
  \BetaN+\frac{\SINRthresholdGUE}{\muG}\right)\noise\right)\Bigg]
  \Biggl].
\end{align}
\begin{IEEEproof}
The rate coverage probability $P_4$ can be expressed as
\begin{subequations}\label{eq:P4_proof}
\begin{align}
\nonumber
  P_{4}&=\mathbb{P}_{\AUERecPow,\GUERecPow}\left(\frac{\AUERecPow}{\noise}
  <\SINRthresholdAUE, \frac{\GUERecPow}{\AUERecPow+\noise}\geq
  \SINRthresholdGUE, \frac{\AUERecPow}{\GUERecPow+\noise}<\SINRthresholdAUE \right)\\ \label{eq:P4_proof_3ineq}
  &=\mathbb{P}_{\AUERecPow,\GUERecPow}\left(\AUERecPow<\SINRthresholdAUE\noise
  ,\GUERecPow\geq\SINRthresholdGUE(\AUERecPow+\noise),\AUERecPow<
  \SINRthresholdAUE(\GUERecPow+\noise)\right)\\ \label{eq:P4_proof_2ineq}
  &=\mathbb{P}_{\AUERecPow,\GUERecPow}\left(\AUERecPow<\SINRthresholdAUE
  \noise, \GUERecPow\geq\SINRthresholdGUE(\AUERecPow+\noise)\right)
  =\mathbb{E}_{\AUERecPow}\left[\mathbb{P}_{\GUERecPow}\left(
  a<\SINRthresholdAUE\noise, \GUERecPow\geq\SINRthresholdGUE(a+\noise)\right)\right]\\ \nonumber
  &=\int_{0}^{\SINRthresholdAUE\noise}\mathbb{P}_{\GUERecPow}\left(
  \GUERecPow\geq\SINRthresholdGUE(a+\noise)\right) f_{\AUERecPow}(a)da \\ \label{eq:P4_proof_intProb}
  &=\int_{0}^{\SINRthresholdAUE\noise}\exp\left(-\frac{\SINRthresholdGUE(a
  +\noise)}{\GUEavgRpow\GUEantenna}\right) f_{\AUERecPow}(a)da.
\end{align}
\noindent The inequalities in \eqref{eq:P4_proof_3ineq} are plotted in Fig. \ref{fig:P3_P4_plots}. Note that the areas enclosed by the relevant curves do not vary with the value of $\SINRthresholdAUE\SINRthresholdGUE$. \eqref{eq:P4_proof_2ineq} is the simplified expression for the quadrilateral area of $P_{4}$. In \eqref{eq:P4_proof_intProb}, the evaluation of $\mathbb{P}_{\GUERecPow}\left(\GUERecPow\geq\SINRthresholdAUE(a+\noise)\right)$ is similar to that in \eqref{eq:P3_proofG_intProb}. By integrating \eqref{eq:P4_proof_intProb} with respect to $a$, we obtain \eqref{eq:P4}.
\end{subequations}
\end{IEEEproof}
\end{proposition}

The final rate coverage probabilities $P_{\mathrm{Tot}}$, $P_{\mathrm{AUE}}$ and $P_{\mathrm{TUE}}$ can be computed after evaluating $P_1$, $P_2$, $P_3$ and $P_4$, as summarized in Fig. \ref{fig:metric_diagram}.

\section{Numerical Results}\label{sec:results}
In this section, we investigate the performance of the proposed NOMA scheme with respect to the AUE SINR threshold, TUE SINR threshold and AUE altitude. The accuracy of the derived analytical expressions is validated by comparing them with simulation results. The parameter values used for the results are given in Table \ref{table:parameter_val}. The chosen values are consistent with other relevant works in the literature \cite{zhou2018underlay,3gpptr36.777,fotouhi2019survey}. We assume a bandwidth of $10$ MHz and consider AUE target SINR thresholds $\{0,10,20,30,40\}$ dB, corresponding to AUE target rates $\{10,34.6,66.6,99.7,134.6\}$ Mbps. We consider a target rate of $10$ Mbps (corresponding to a target SINR threshold of $0$ dB) for the TUE, unless stated otherwise.

\subsection{Probabilistic LoS model} \label{subsec:LoSmodels}
\begin{table}
\centering
  \caption{Parameter values for numerical and simulation results.}\label{table:parameter_val}
\begin{tabular}{llllll}
\hline
\multicolumn{1}{|l|}{\textbf{Parameter}}                    & \multicolumn{1}{l|}{\textbf{Symbol}} & \multicolumn{1}{l||}{\textbf{Value}} & \multicolumn{1}{l|}{\textbf{Parameter}}                 & \multicolumn{1}{l|}{\textbf{Symbol}} & \multicolumn{1}{l|}{\textbf{Value}} \\ \hline
\multicolumn{1}{|l|}{Cell radius}                           & \multicolumn{1}{l|}{$\cellrad$}               & \multicolumn{1}{l||}{$500$ m}              & \multicolumn{1}{l|}{Number of rounds}                   & \multicolumn{1}{l|}{$\rounds$}               & \multicolumn{1}{l|}{$3$}              \\ \hline
\multicolumn{1}{|l|}{Height of BS}                          & \multicolumn{1}{l|}{$\hBS$}               & \multicolumn{1}{l||}{$30$ m}              & \multicolumn{1}{l|}{Time period for AUE's transmission} & \multicolumn{1}{l|}{$\TPeriod$}               & \multicolumn{1}{l|}{$30$ s}              \\ \hline
\multicolumn{1}{|l|}{Noise power}                           & \multicolumn{1}{l|}{$\noise$}                & \multicolumn{1}{l||}{$-100$ dBm}               & \multicolumn{1}{l|}{AUE's transmit power}               & \multicolumn{1}{l|}{$\AUETranPow$}                & \multicolumn{1}{l|}{$0.1$ W}               \\ \hline
\multicolumn{3}{|c||}{\textbf{TUE}}                                                                                                       & \multicolumn{1}{l|}{AUE's antenna gain}                 & \multicolumn{1}{l|}{$\AUEantenna$}                & \multicolumn{1}{l|}{$1$}               \\ \hline
\multicolumn{1}{|l|}{TUE's cutoff threshold}                & \multicolumn{1}{l|}{$\GUEavgRpow$}                & \multicolumn{1}{l||}{$-75$ dBm}               & \multicolumn{1}{l|}{Pathloss exponent for LoS link}     & \multicolumn{1}{l|}{$\alphaLOS$}                & \multicolumn{1}{l|}{$2.2$}               \\ \hline
\multicolumn{1}{|l|}{TUE's antenna gain}                    & \multicolumn{1}{l|}{$\GUEantenna$}                & \multicolumn{1}{l||}{$1$}               & \multicolumn{1}{l|}{Pathloss exponent for NLoS link}    & \multicolumn{1}{l|}{$\alphaNLOS$}                & \multicolumn{1}{l|}{$3.5$}               \\ \hline
\multicolumn{1}{|l|}{Pathloss exponent of terrestrial link} & \multicolumn{1}{l|}{$\alphaGUE$}                & \multicolumn{1}{l||}{3.5}               & \multicolumn{1}{l|}{Attenuation for LoS link}           & \multicolumn{1}{l|}{$\etaLOS$}                & \multicolumn{1}{l|}{$0$ dB}               \\ \hline
\multicolumn{3}{|c||}{\textbf{AUE}}                                                                                                       & \multicolumn{1}{l|}{Attenuation for NLoS link}          & \multicolumn{1}{l|}{$\etaNLOS$}                & \multicolumn{1}{l|}{$13$ dB}               \\ \hline
\multicolumn{1}{|l|}{AUE's speed}                           & \multicolumn{1}{l|}{$\AUEspeed$}                & \multicolumn{1}{l||}{$15$ m/s}               & \multicolumn{1}{l|}{Fading parameter for LoS link}      & \multicolumn{1}{l|}{$\mLOS$}                & \multicolumn{1}{l|}{5}               \\ \hline
\multicolumn{1}{|l|}{AUE's altitude}                        & \multicolumn{1}{l|}{$\hAUE$}                & \multicolumn{1}{l||}{$25$ m, $120$ m}               & \multicolumn{1}{l|}{Fading parameter for NLoS link}     & \multicolumn{1}{l|}{$\mNLOS$}                & \multicolumn{1}{l|}{$1$}               \\ \hline
\end{tabular}
\end{table}

We adopt the probabilistic LoS model suggested in ITU recommendation report \cite{itu2012propagation} by considering AUE as the transmitter and the terrestrial BS as the receiver. Thus, the probability of LoS between the AUE and the terrestrial BS is \cite{cherif2019downlink}
\begin{align}\label{eq:LoS_itu}
  \PLOS=\prod_{n_{\mathrm{ITU}=0}}^{m_{\mathrm{ITU}}}\left[
  1-\exp\left(-\frac{{\left[\hAUE-\frac{\left(n_{\mathrm{ITU}}+\frac{1}{2}\right)
  \left(\hAUE-\hBS\right)}{m_{\mathrm{ITU}}+1}\right]}^2}{2\delta_{\mathrm{ITU}}}
  \right)\right],
\end{align}
\noindent where $m_{\mathrm{ITU}}=\left\lfloor\left(\frac{\dAUEhori\sqrt{\alpha_{\mathrm{ITU}}
\beta_{\mathrm{ITU}}}}{1000}-1\right)\right\rfloor$ and $\lfloor\cdot\rfloor$ is the floor function. $\alpha_{\mathrm{ITU}}$, $\beta_{\mathrm{ITU}}$ and $\delta_{\mathrm{ITU}}$ correspond to the environment-related parameters that can be used to describe the built-up area. $\alpha_{\mathrm{ITU}}$ and $\beta_{\mathrm{ITU}}$ correspond to the ratio of land area covered by buildings to total land area, and the average number of buildings per unit area, respectively. $\delta_{\mathrm{ITU}}$ is the parameter for the Rayleigh distribution that determines the building heights. The values of these parameters for suburban, urban, dense urban and urban high-rise environments are given in Table I in Page $2$ in \cite{holis2008elevation}.

\begin{figure}[t]
\centering
\subfigure[]{\label{fig:PLOS}\includegraphics[width=0.48\textwidth]{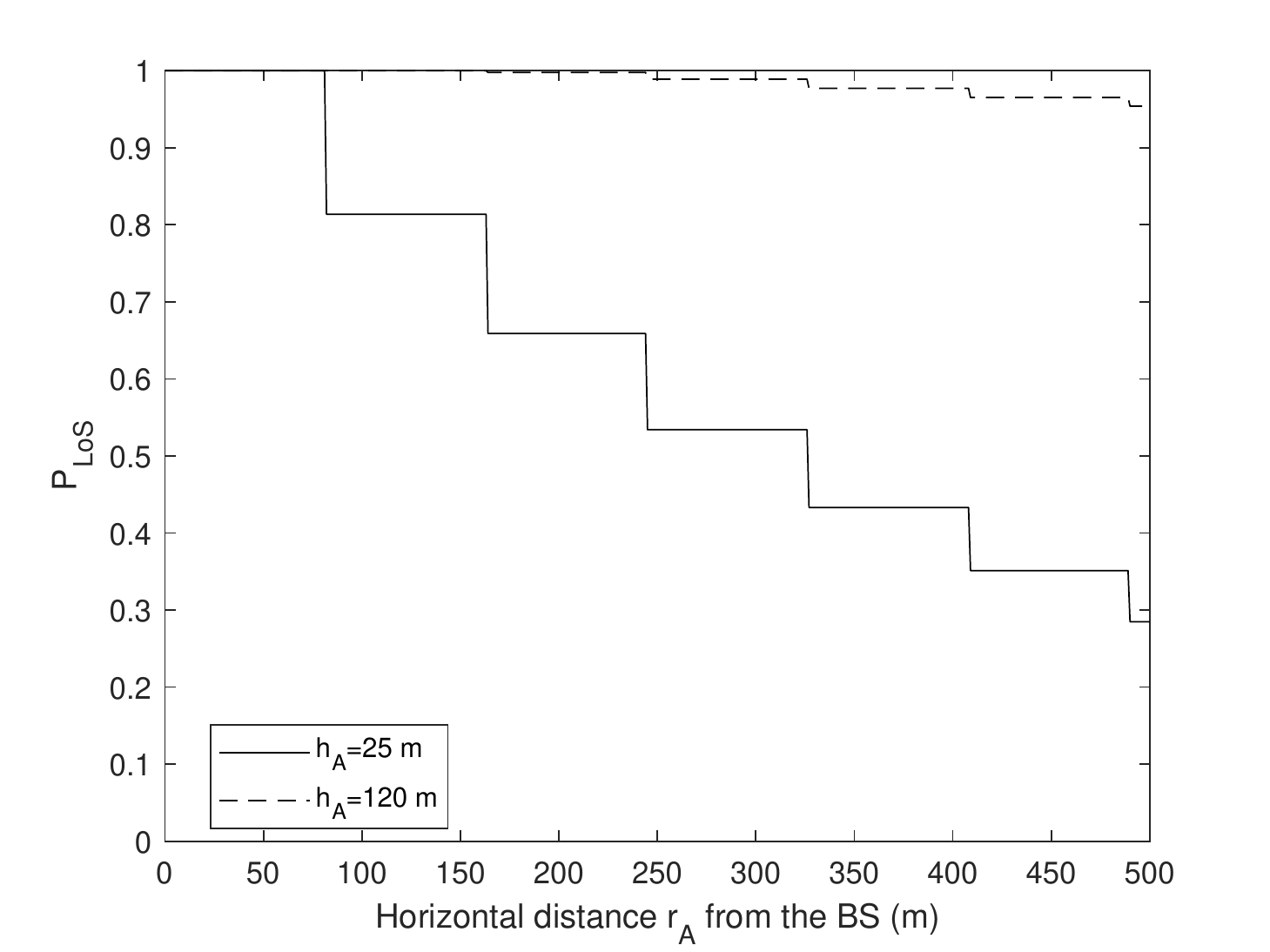}}
\subfigure[]{\label{fig:PLOS_elevation}\includegraphics[width=0.48\textwidth]{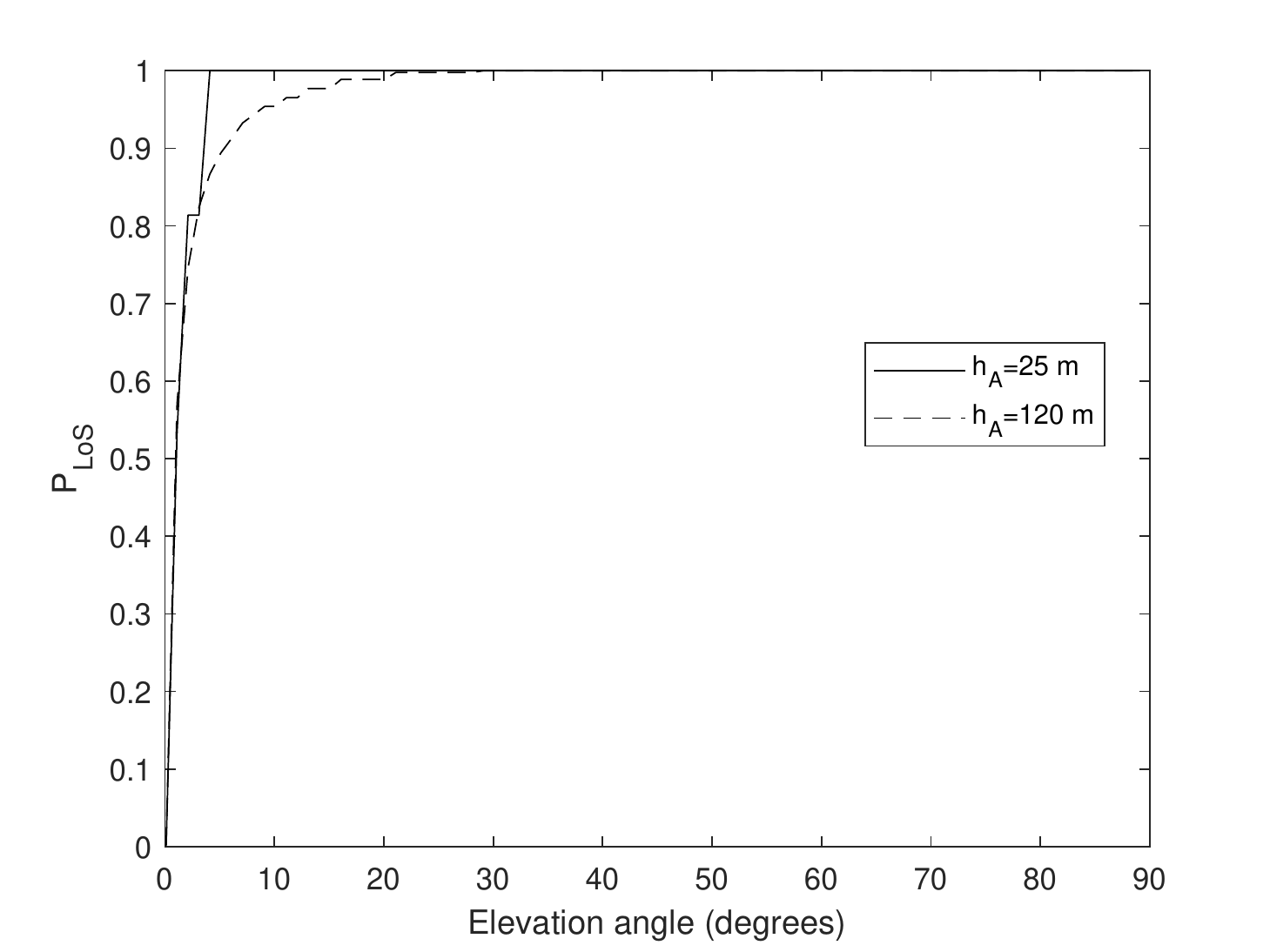}}
\caption{Probability of line-of-sight, $\PLOS$ versus (a) horizontal distance from the BS $\dAUEhori$ (m), and (b) elevation angle between the AUE and BS. The adopted values for the environmental parameters corresponding to the urban built-up area are, $\alpha_{\mathrm{ITU}}=0.3$, $\beta_{\mathrm{ITU}}=500$ and $\delta_{\mathrm{ITU}}=15$. }\label{fig:PLOS_models}
\end{figure}

Fig. \ref{fig:PLOS_models} illustrates the probabilistic LoS behavior for two different AUE altitudes. Fig. \ref{fig:PLOS} demonstrates the variation of LoS probability with the horizontal distance of AUE from the BS, whereas Fig. \ref{fig:PLOS_elevation} shows the variation of the LoS probability with angle of elevation between the AUE and BS. Note that both plots exhibit step-wise discrete behavior due to the blockage caused by buildings in the urban built-up area. This behavior becomes smooth and continuous at very high AUE altitudes. However, in this work, we consider a practical AUE altitude range of $25-300$ m.

\subsection{AUE Trajectory Model}
For the purpose of generating the results, we model the trajectory of AUE using an Archimedes' spiral. The Archimedes' spiral has a special property that any ray from the origin intersects successive turnings of the spiral in points with a constant separation distance. Hence, it is a suitable trajectory for monitoring or surveillance in a disk region. \textit{Note that the proposed framework in this paper is valid for any trajectory model}. In Sections \ref{sec:results} C-E we consider the Archimedes' spiral trajectory and in Section \ref{subsec:3gpp} we consider the 3GPP trajectory model.

The AUE starts its spiral trajectory at the center of the cell at a height $\hAUE$. The trajectory of the AUE can be described by a spiral with equation $\dAUEhori=\frac{R}{2\pi\rounds}\AUEazi$, where $\rounds$ is the number of rounds and $\AUEazi$ is the orientation of AUE in the azimuth plane, measured with respect to the $+$x-axis. The equation for $\dAUEhori$ is derived based on the assumption that the spiral starts at the center of the cell and reaches the cell edge at ${\AUEazi}_\mathrm{Edge}=2\pi \rounds$ which is the maximum angle in the azimuth plane for a given number of rounds. Hence, the expressions for $\NPoints$ and $\ndAUEhori$ for the Archimedes' spiral are given as follows:

The number of transmission points along the AUE trajectory defined by the Archimedes' spiral is given by
\begin{align}\label{eq:N_points}
  \NPoints=\left\lfloor\frac{\cellrad\left(2\pi\rounds\sqrt{1+{(2\pi\rounds)}^2}+
  \sinh^{-1}(2\pi\rounds)
  \right)}{4\pi\rounds\AUEspeed\TPeriod}\right\rfloor.
\end{align}

The horizontal distance $\ndAUEhori$ between the AUE and BS at the $n$th transmission point is
\begin{align}\label{eq:rA_n}
  \ndAUEhori=\cellrad\sqrt{\frac{n}{\NPoints}},
\end{align}
\noindent where $n=1,\ldots,\NPoints$.

\begin{figure}[t]
  \centering
  \includegraphics[width=0.4\linewidth]{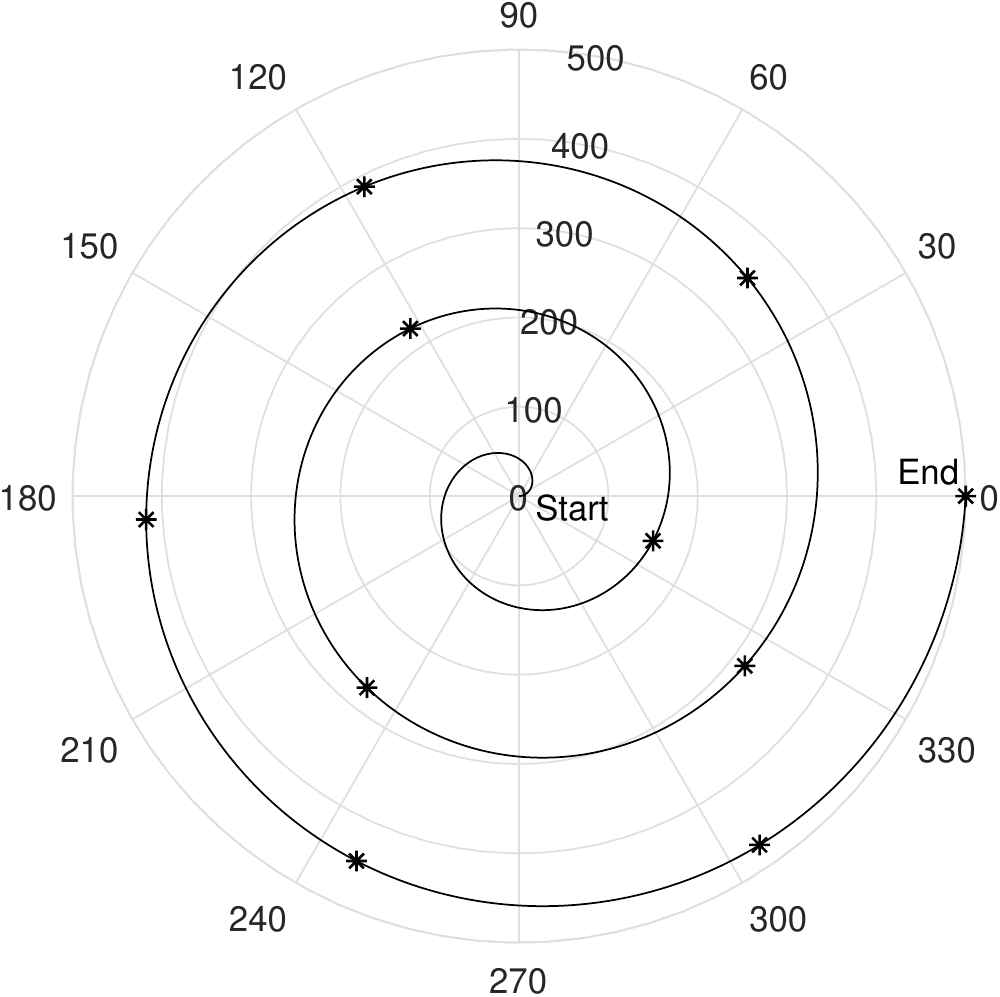}
  \caption{Archimedes' spiral trajectory with $\rounds=3$ and $R=500$ m. AUE's transmission points are denoted by asterisk ($\AUEspeed=15$ m$/$s, $\TPeriod=30$ s).}
  \label{fig:trajectory}
\end{figure}

Fig. \ref{fig:trajectory} demonstrates the Archimedes' trajectory for the case where the AUE travels three complete rounds to reach the cell edge. The AUE performs uplink transmission at 10 points along the trajectory, and the final transmission point lies at the cell edge.

\subsection{Model Validation}\label{subsec:model_valid}
\begin{figure}[t]
\centering
\subfigure[$P_{\mathrm{Tot}}$ at $\hAUE$=$25$ m ]{\label{fig:PT_25_valid}\includegraphics[width=0.44\textwidth]{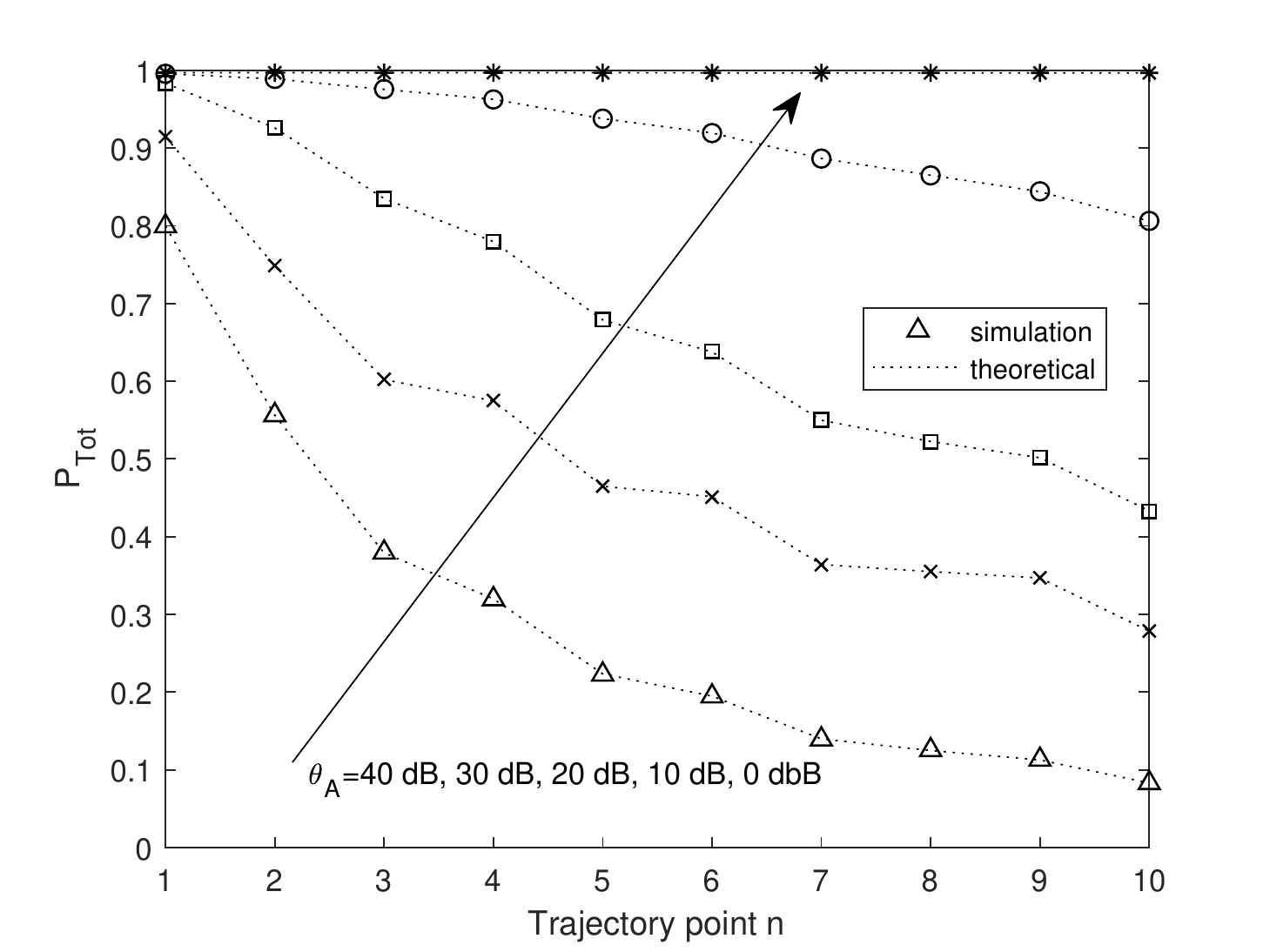}}
\subfigure[$P_{\mathrm{Tot}}$ at $\hAUE$=$120$ m ]{\label{fig:PT_120_valid}\includegraphics[width=0.44\textwidth]{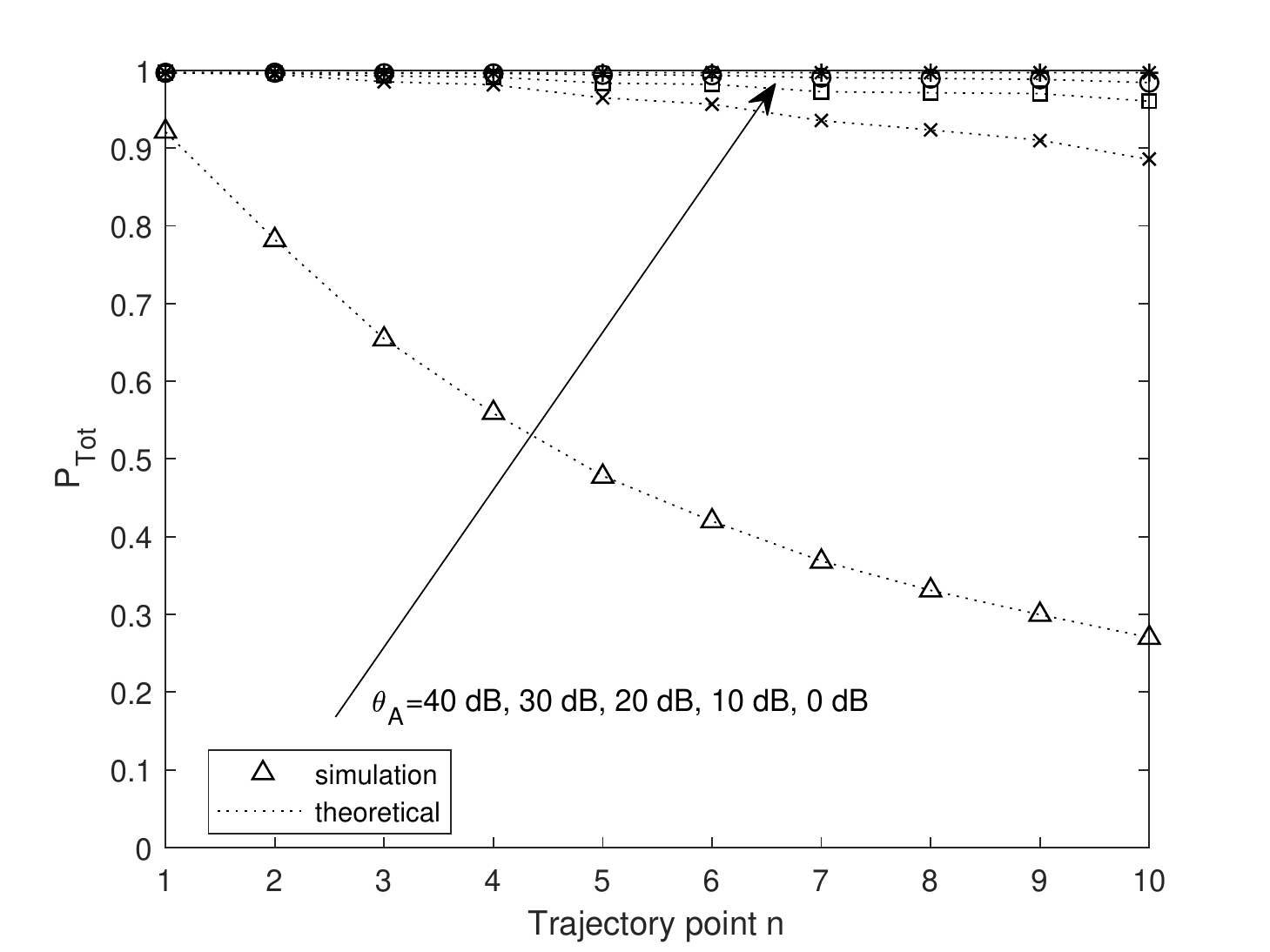}}
\caption{Rate coverage probabilities of the events where both AUE and TUE are decoded ($P_{\mathrm{Tot}}$) for (a) $\hAUE=25$ m and (b) $\hAUE=120$ m. The simulation values and the theoretical values are represented by markers and dotted lines, respectively.}\label{fig:valid}
\end{figure}

Fig. \ref{fig:valid} presents the total rate coverage probability $P_{\mathrm{Tot}}$  for an aerial-terrestrial network where the AUE flies according to an Archimedes' trajectory at a fixed height $\hAUE$ along the entire trajectory. The model validation results are presented only for $P_{\mathrm{Tot}}$, since $P_{\mathrm{AUE}}$ and $P_{\mathrm{TUE}}$ exhibit similar trends as that for $P_{\mathrm{Tot}}$. We present the results when $\SINRthresholdGUE=0$ dB and AUE flies at altitudes $25$ m and $120$ m. We see that the analytical results match well with the simulation results. This validates the accuracy of the analytical framework in Section \ref{sec:analytical}. The figure shows that $P_{\mathrm{Tot}}$ generally decreases when AUE's SINR threshold (corresponds to AUE's target data rate) increases and when the horizontal distance between the AUE and the BS increases, for both altitudes. For the ITU probabilistic LoS model, the performance is better at $120$ m compared to that at $25$ m, and the rate coverage probabilities decrease when the AUE flies away from the BS. This is due to the fact that in the ITU probabilistic LoS model, the probability of LoS is higher at $120$ m compared to that in $25$ m and, the probability of LoS decreases with the increase in $\dAUEhori$ (see Fig. \ref{fig:PLOS}).

In the next two subsections, we only present the numerical results using the proposed analytical framework due to the accuracy of our analytical results.

\subsection{Impact of SINR Thresholds of TUE and AUE}\label{subsec:SINRthreshold}
\begin{figure}[t]
\centering
\subfigure[$\SINRthresholdGUE=0$ dB and $\hAUE=25$ m ]{\label{fig:gue0_hA25}\includegraphics[width=0.44\textwidth]{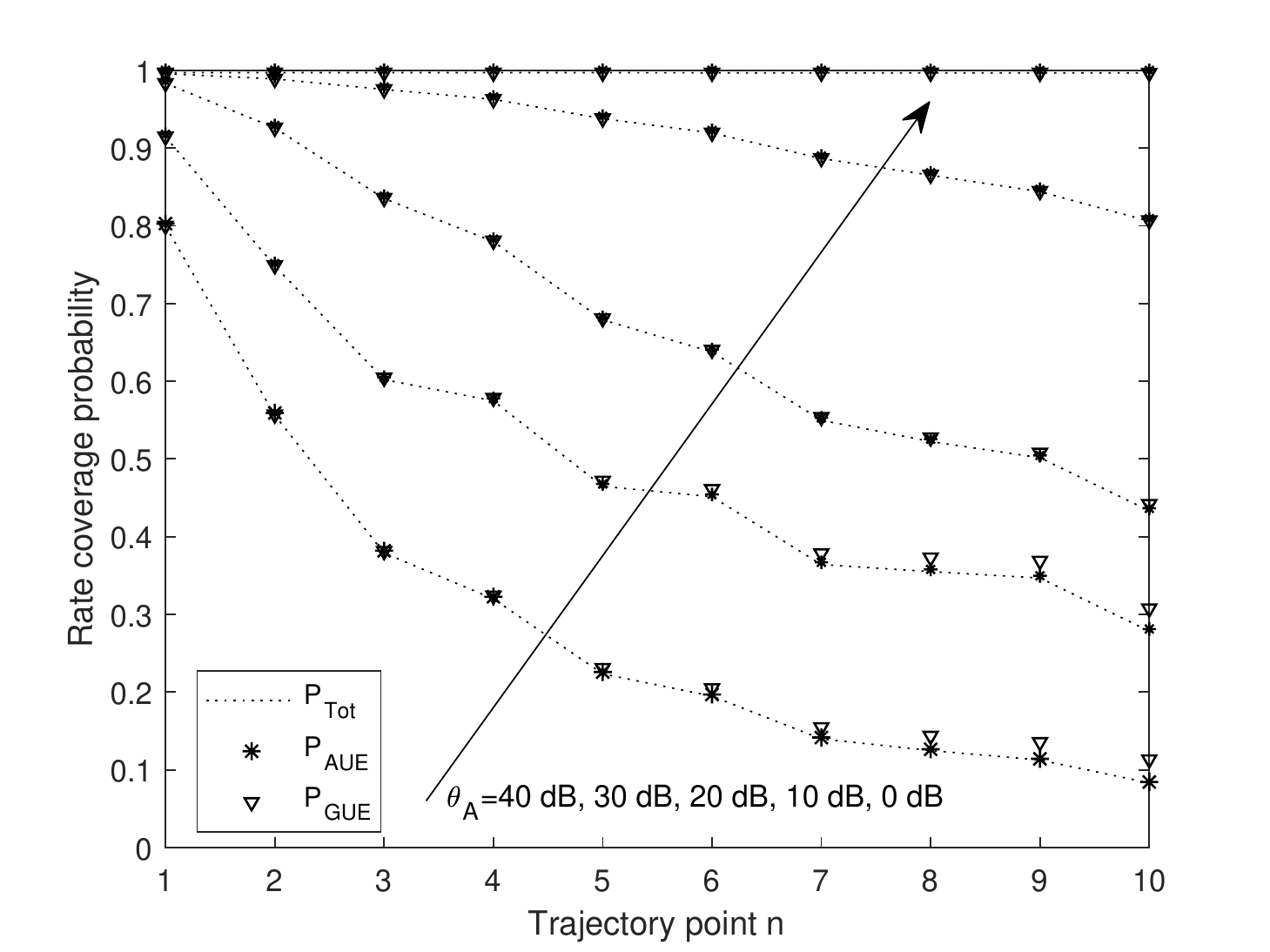}}
\subfigure[$\SINRthresholdGUE=10$ dB and $\hAUE=25$ m ]{\label{fig:gue10_hA25}\includegraphics[width=0.44\textwidth]{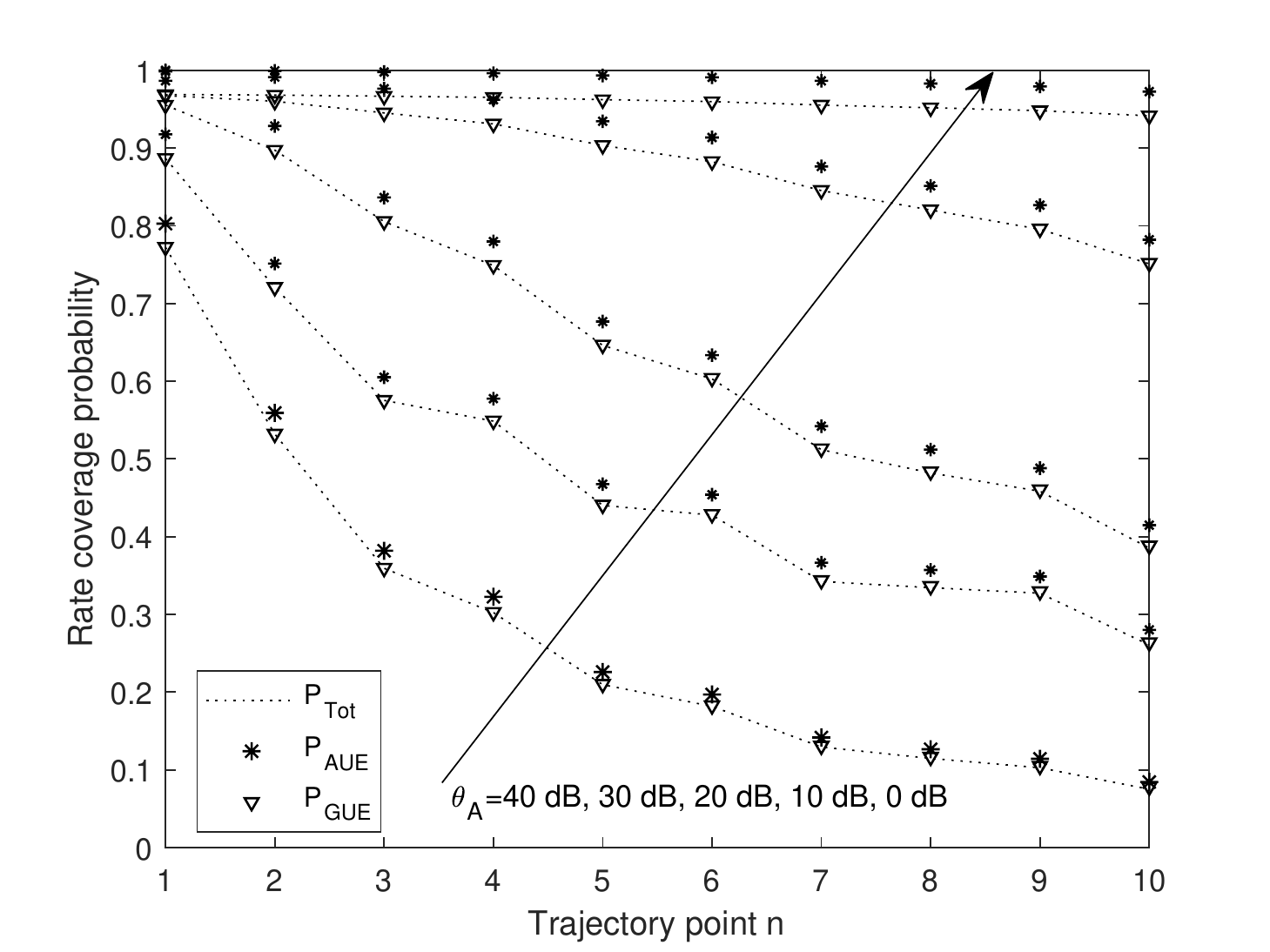}}\\
\subfigure[$\SINRthresholdGUE=0$ dB and $\hAUE=120$ m ]{\label{fig:gue0_hA120}\includegraphics[width=0.44\textwidth]{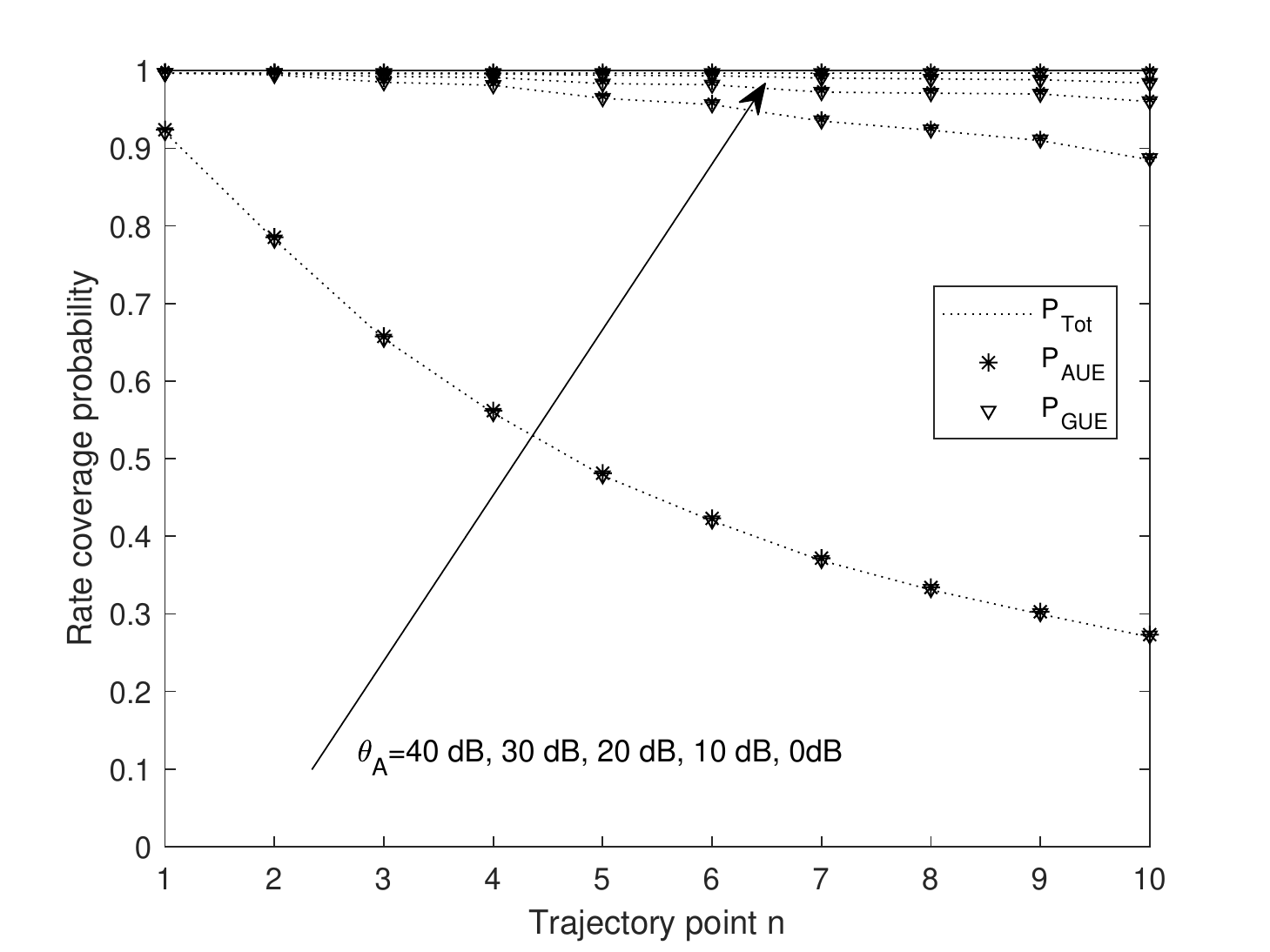}}
\subfigure[$\SINRthresholdGUE=10$ dB and $\hAUE=120$ m ]{\label{fig:gue10_hA120}\includegraphics[width=0.44\textwidth]{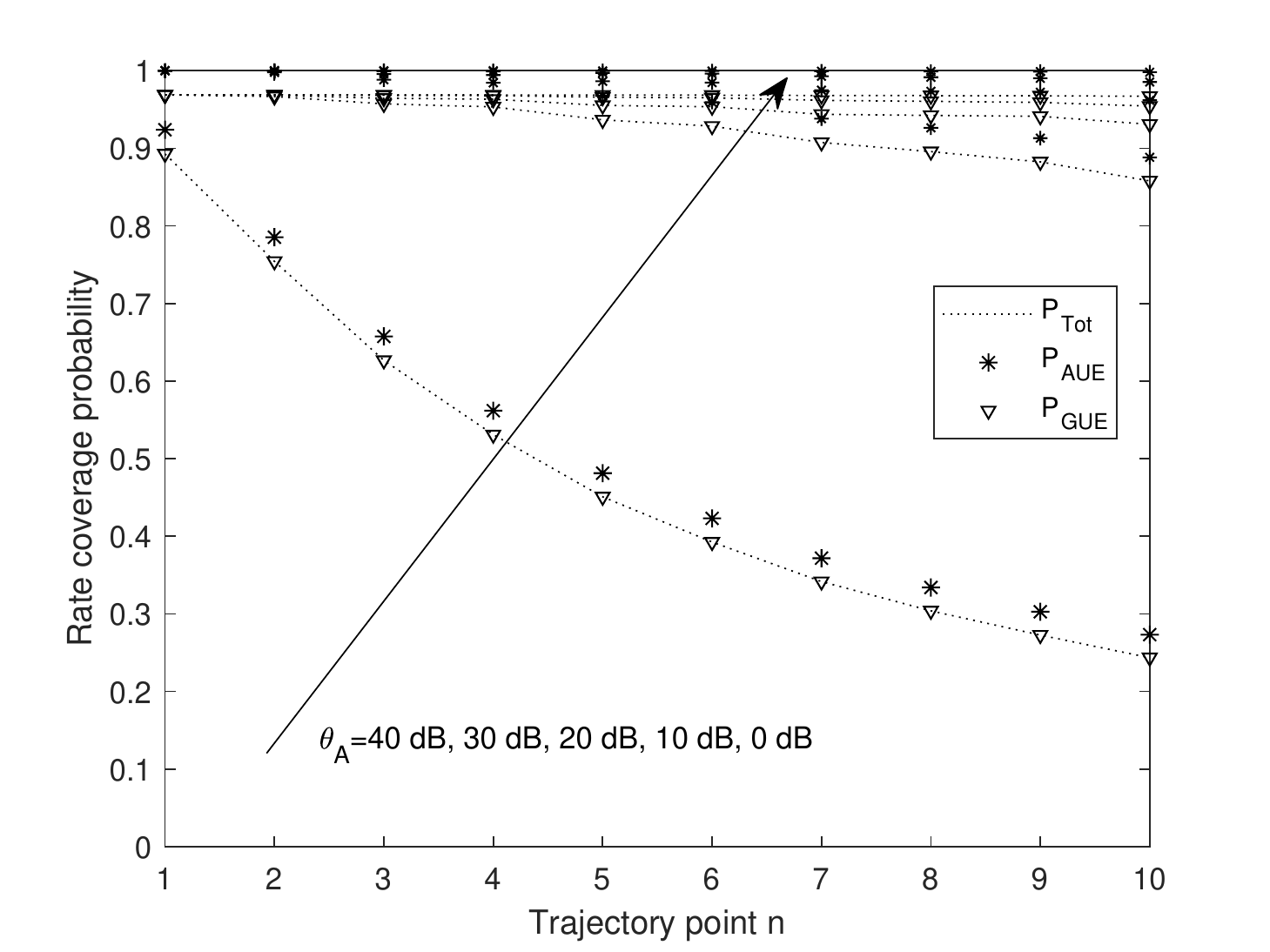}}\\
\caption{Rate coverage probabilities $P_{\mathrm{Tot}}$, $P_{\mathrm{AUE}}$ and $P_{\mathrm{TUE}}$ vs. trajectory point $n$ for different $\SINRthresholdAUE=0$, $10$, $20$, $30$, $40$ dB and, (a) $\SINRthresholdGUE=0$ dB and $\hAUE=25$ m, (b) $\SINRthresholdGUE=10$ dB and $\hAUE=25$ m, (c) $\SINRthresholdGUE=0$ dB and $\hAUE=120$ m and (d) $\SINRthresholdGUE=10$ dB and $\hAUE=120$ m. $P_{\mathrm{Tot}}$, $P_{\mathrm{AUE}}$ and $P_{\mathrm{TUE}}$ are denoted by solid line, asterisk and circle, respectively.}\label{fig:impactGUE}
\end{figure}

Fig. \ref{fig:impactGUE} shows the impact of $\SINRthresholdGUE$ (equivalently, the target data rate of TUE), on rate coverage probabilities $P_{\mathrm{Tot}}$, $P_{\mathrm{AUE}}$ and $P_{\mathrm{TUE}}$ for different $\SINRthresholdAUE$ values along the spiral trajectory. The results are presented for two different AUE heights ($25$ m and $120$ m). $n=1$ corresponds to the first transmission point and $n=10$ corresponds to the last transmission point when the AUE reaches the cell edge (see transmission points marked in Fig. \ref{fig:trajectory}).

Fig. \ref{fig:impactGUE} shows that depending upon the SINR threshold of TUE and SINR threshold of AUE, $P_{\mathrm{Tot}}$ can be dominated by either decoding of TUE or AUE. For both heights, when $\SINRthresholdGUE=0$ dB (see Fig. \ref{fig:gue0_hA25} and \ref{fig:gue0_hA120}) we can see that $P_{\mathrm{Tot}}$, $P_{\mathrm{AUE}}$ and $P_{\mathrm{TUE}}$ have similar rate coverage probability values for lower $\SINRthresholdAUE$ values ($0-20$ dB). However, this trend changes at $\SINRthresholdAUE=30$ dB and $40$ dB. When AUE is flying at $25$ m (see Fig. \ref{fig:gue0_hA25}) and is closer to the cell edge, $P_{\mathrm{TUE}}$ is slightly higher than $P_{\mathrm{Tot}}$ and $P_{\mathrm{AUE}}$ at $\SINRthresholdAUE=30,40$ dB. This is caused by the high path loss and poor LoS aerial links in proximity to the cell edge. \textit{In this case, $P_{\mathrm{Tot}}$ is dominated by the decoding of AUE}. However, this behavior becomes less prominent at higher AUE heights (see Fig. \ref{fig:gue0_hA120}). This is because at higher AUE heights, the AUE is almost always guaranteed to be successfully decoded due to the strong LoS aerial links.

Fig. \ref{fig:gue10_hA25} and \ref{fig:gue10_hA120} compare the rate coverage probabilities at $\SINRthresholdGUE=10$ dB for $\hAUE=25$ and $120$ m. At both heights, $P_{\mathrm{Tot}}$ is similar to $P_{\mathrm{TUE}}$ for low to moderate $\SINRthresholdAUE$ values ($0$, $10$, $20$, $30$ dB). This implies that \textit{$P_{\mathrm{Tot}}$ is dominated by the decoding of TUE} at higher $\SINRthresholdGUE$ for low to moderate $\SINRthresholdAUE$. In Fig. \ref{fig:gue10_hA25}, we can see that this effect becomes insignificant when $\SINRthresholdAUE$ is higher and AUE is flying at a relatively low height. It is important to note that this behavior in Fig. \ref{fig:gue10_hA25} is consistent throughout the trajectory at a higher $\SINRthresholdGUE$, unlike in Fig. \ref{fig:gue0_hA25}, where the trend changes as the AUE moves from cell center to cell edge. At $120$ m, due to the strong LoS aerial links, $P_{\mathrm{Tot}}$ is dominated by the decoding of TUE for all considered $\SINRthresholdAUE$ values throughout the trajectory at higher $\SINRthresholdGUE$ values (see Fig. \ref{fig:gue10_hA120}). Note that the trends of $P_{\mathrm{AUE}}$ and $P_{\mathrm{TUE}}$ are similar to that of $P_{\mathrm{Tot}}$ for each of the cases presented in Fig. \ref{fig:impactGUE}. Thus, for the rest of the results section, we only present the corresponding results for $P_{\mathrm{Tot}}$.

\subsection{Impact of AUE Altitude and Built-Up Environments}\label{subsec:AUEheight}
\begin{figure}[t]
\centering
\subfigure[suburban ]{\label{fig:suburban}\includegraphics[width=0.44\textwidth]{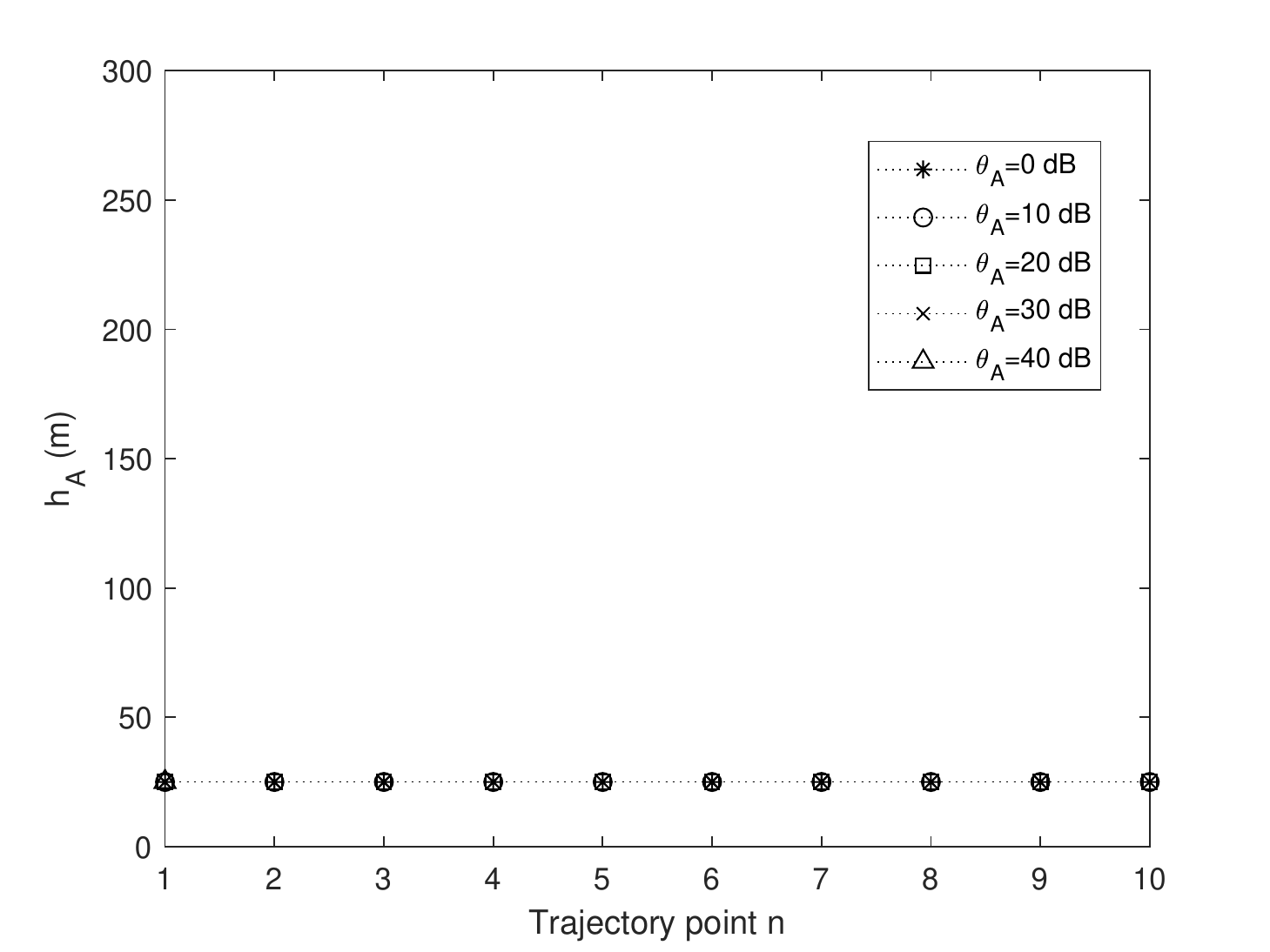}}
\subfigure[urban ]{\label{fig:urban}\includegraphics[width=0.44\textwidth]{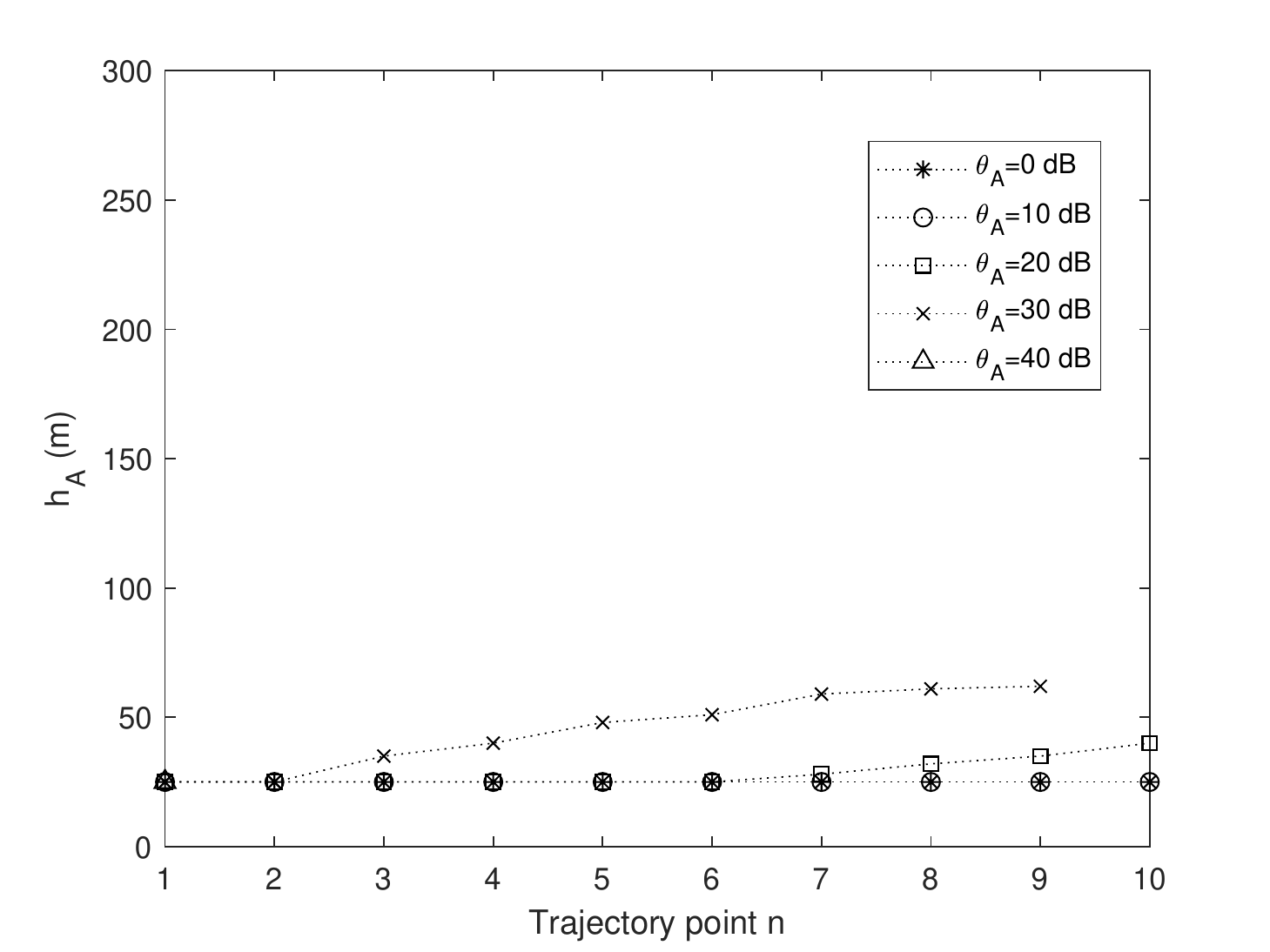}}\\
\subfigure[dense urban ]{\label{fig:dense}\includegraphics[width=0.44\textwidth]{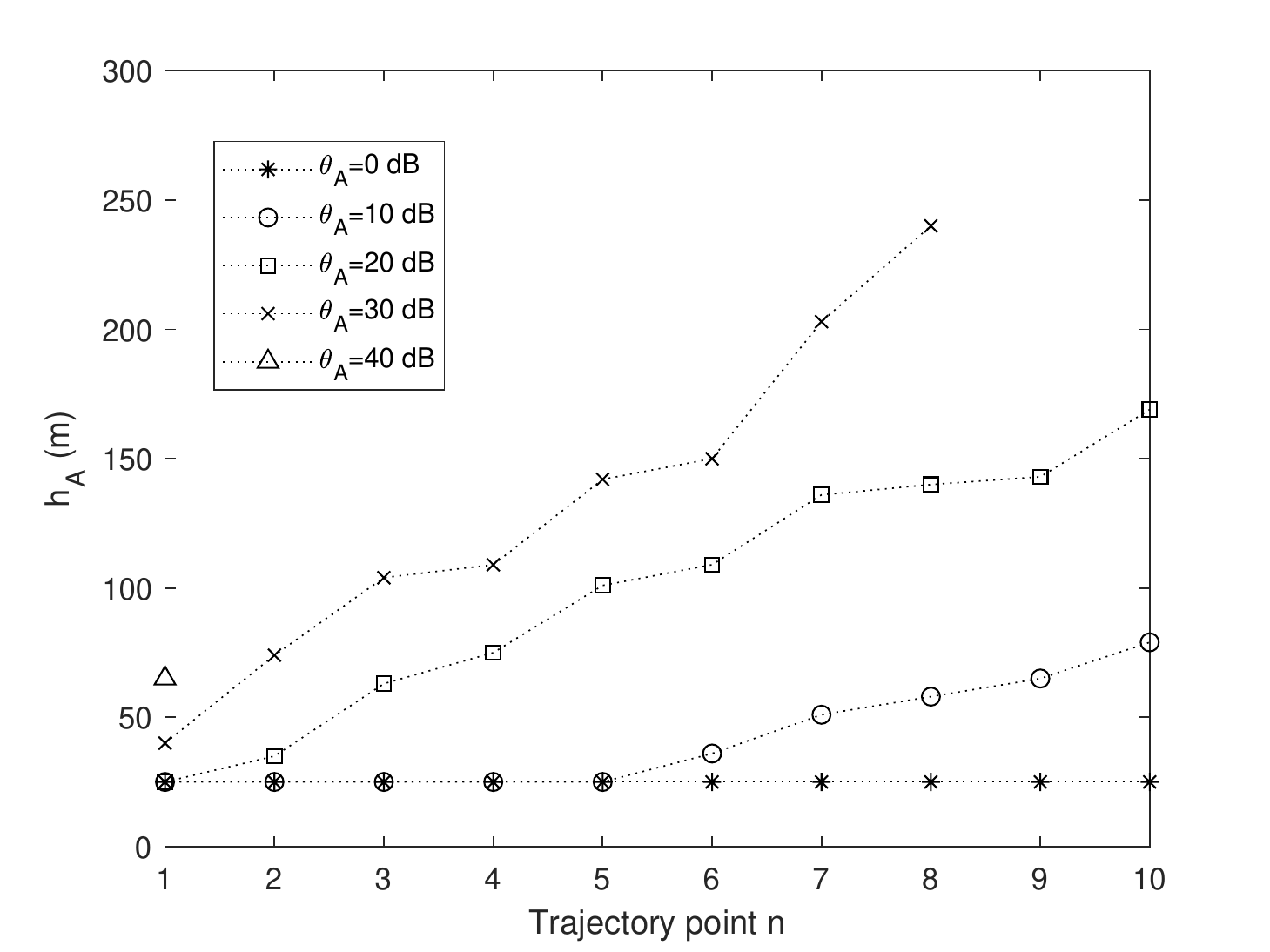}}
\subfigure[urban high-rise ]{\label{fig:highrise}\includegraphics[width=0.44\textwidth]{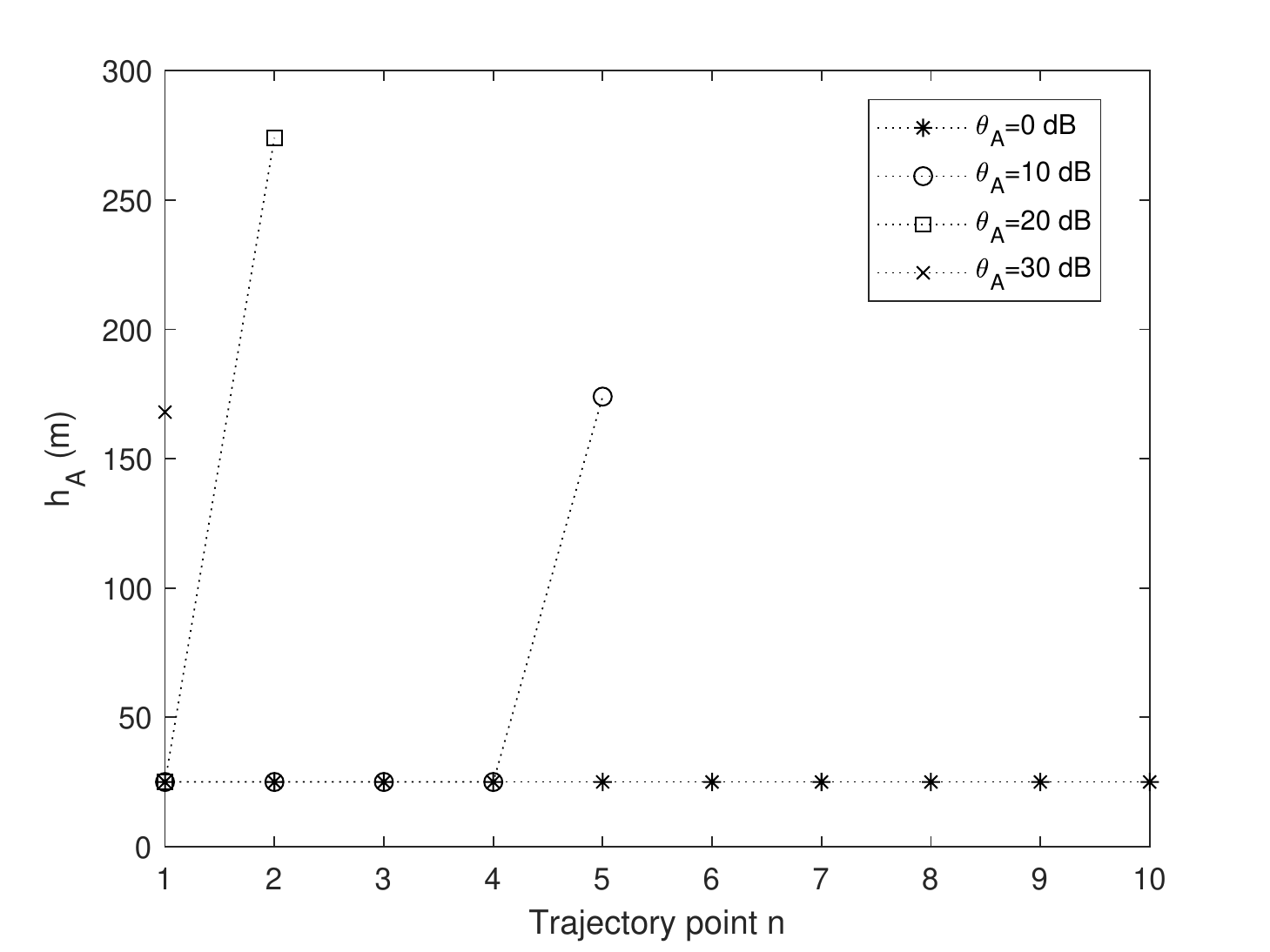}}\\
\caption{Minimum height of AUE to achieve a total rate coverage probability of 0.9 vs. the trajectory point for (a) suburban, (b) urban, (c) dense urban and (d) urban high-rise environments. }\label{fig:env}
\end{figure}

Previously, we assumed that the AUE flies at a constant height in a spiral trajectory. Now we consider the case where the AUE still follows the spiral trajectory, but it can ascend or descend at each trajectory point to achieve a certain quality of service (QoS).

We define the QoS as the probability where both AUE and TUE are decoded, which is equivalent to $P_{\mathrm{Tot}}$. Current regulations in most countries do not permit AUEs to fly higher than a certain height \cite{fotouhi2019survey,lin2018sky}. Hence, in this section, we focus on the minimum height of AUE at each trajectory point, to achieve a QoS of $90\%$ (corresponds to $P_{\mathrm{Tot}}=0.9$) for different built-up environments.

\begin{figure}[t]
\centering
\subfigure[$\SINRthresholdAUE=20$ dB and $n=1$. ]{\label{fig:optH_20n1}\includegraphics[width=0.44\textwidth]{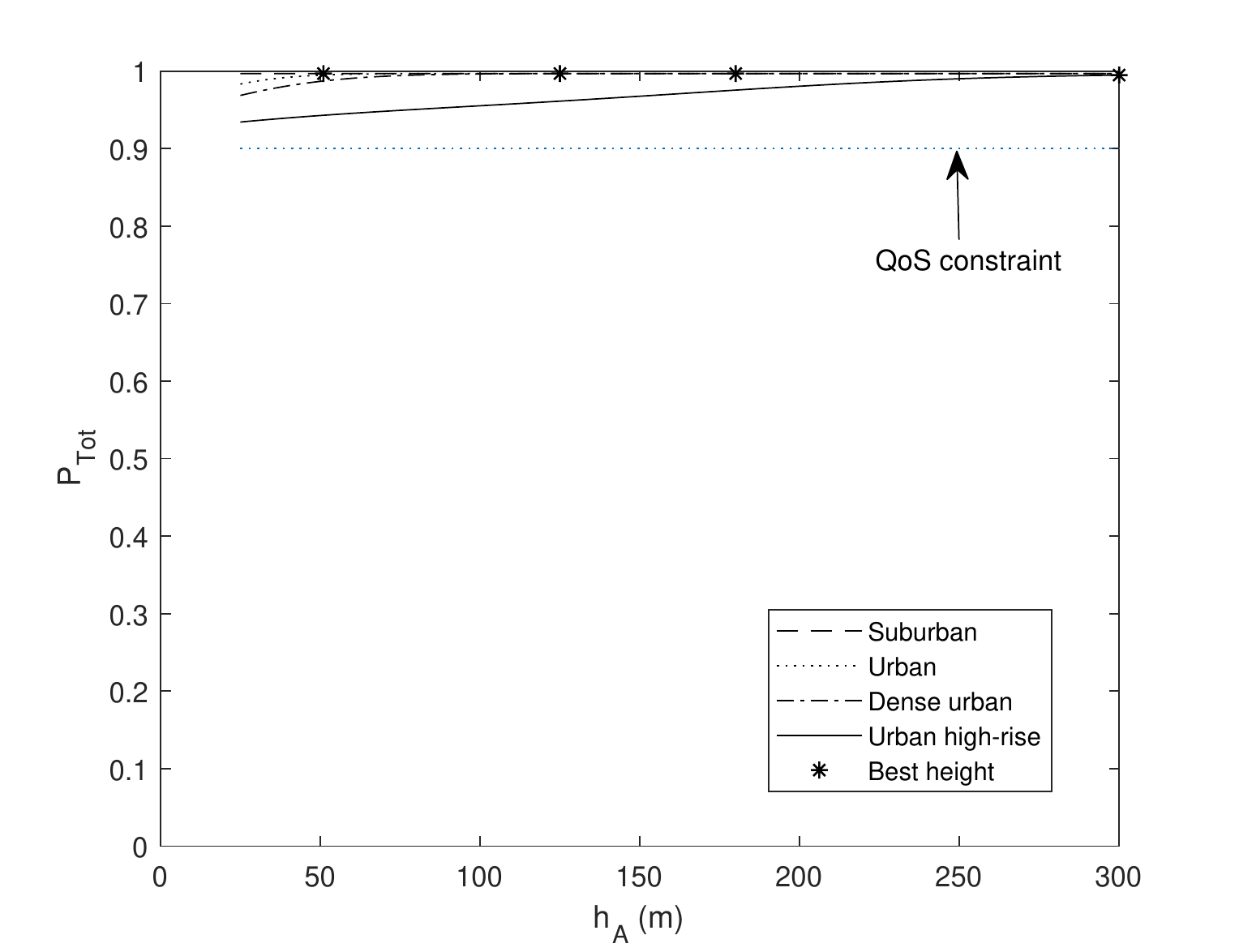}}
\subfigure[$\SINRthresholdAUE=40$ dB and $n=1$ ]{\label{fig:optH_40n1}\includegraphics[width=0.44\textwidth]{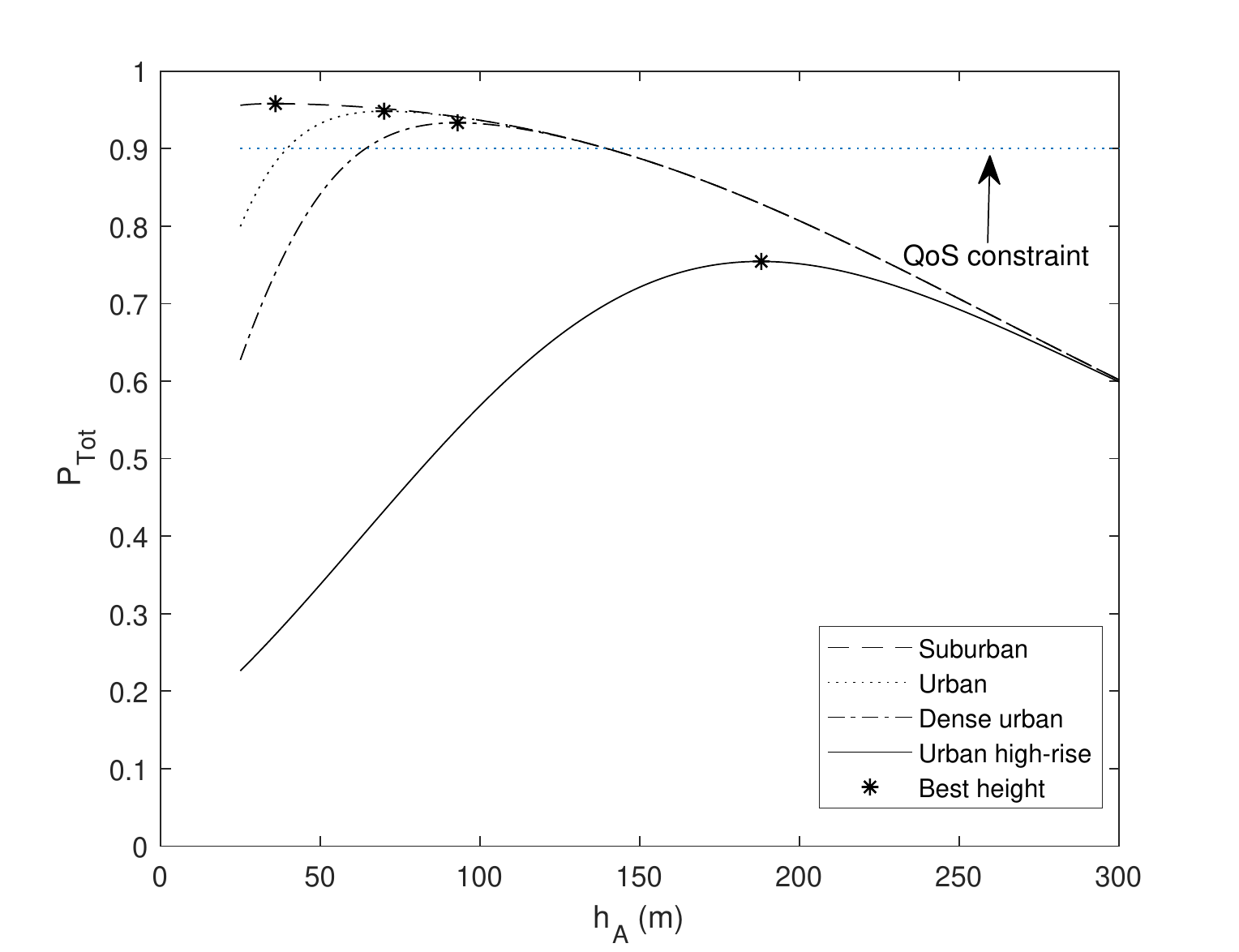}}\\
\subfigure[$\SINRthresholdAUE=20$ dB and $n=10$ ]{\label{fig:optH_20n10}\includegraphics[width=0.44\textwidth]{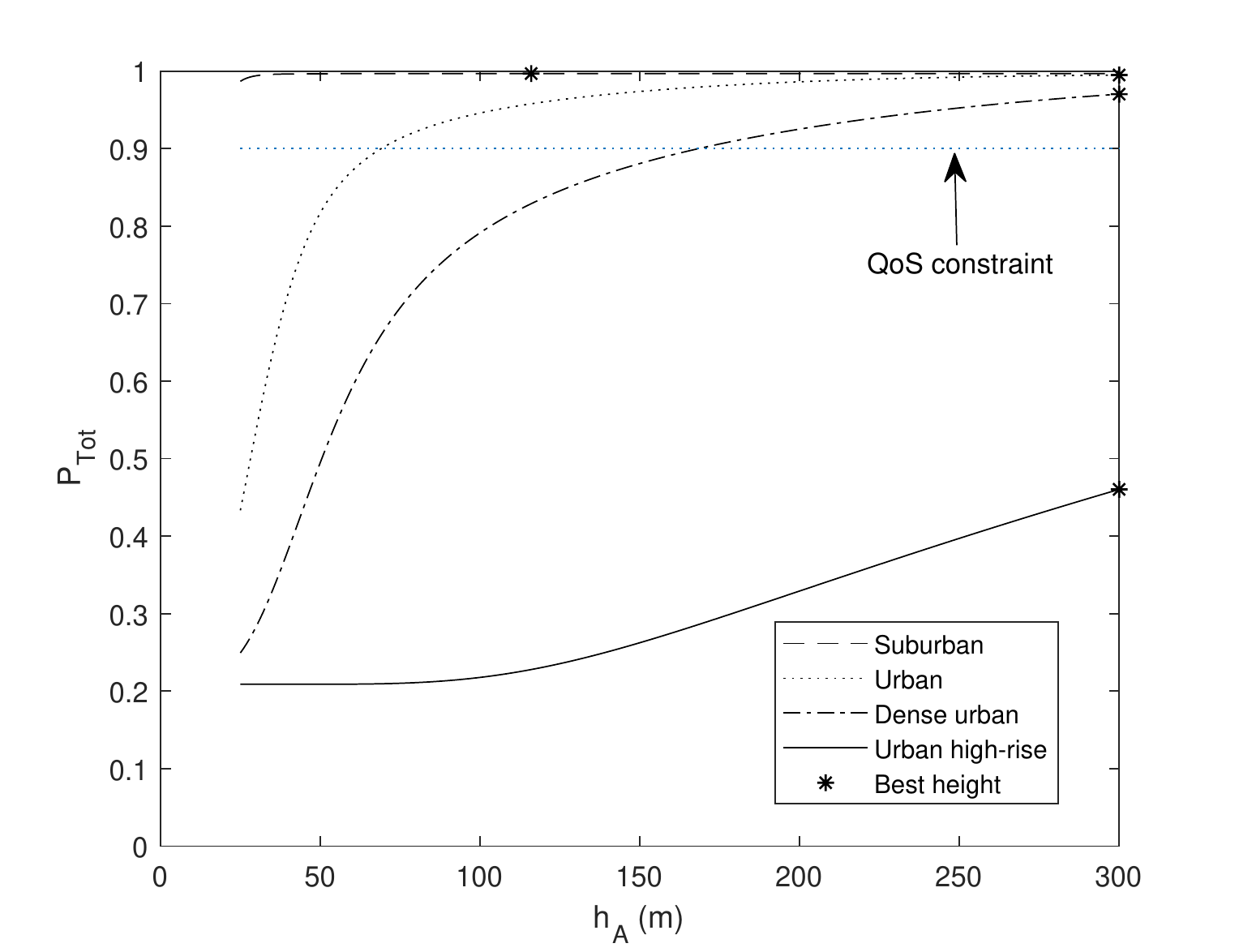}}
\subfigure[$\SINRthresholdAUE=40$ dB and $n=10$ ]{\label{fig:optH_40n10}\includegraphics[width=0.44\textwidth]{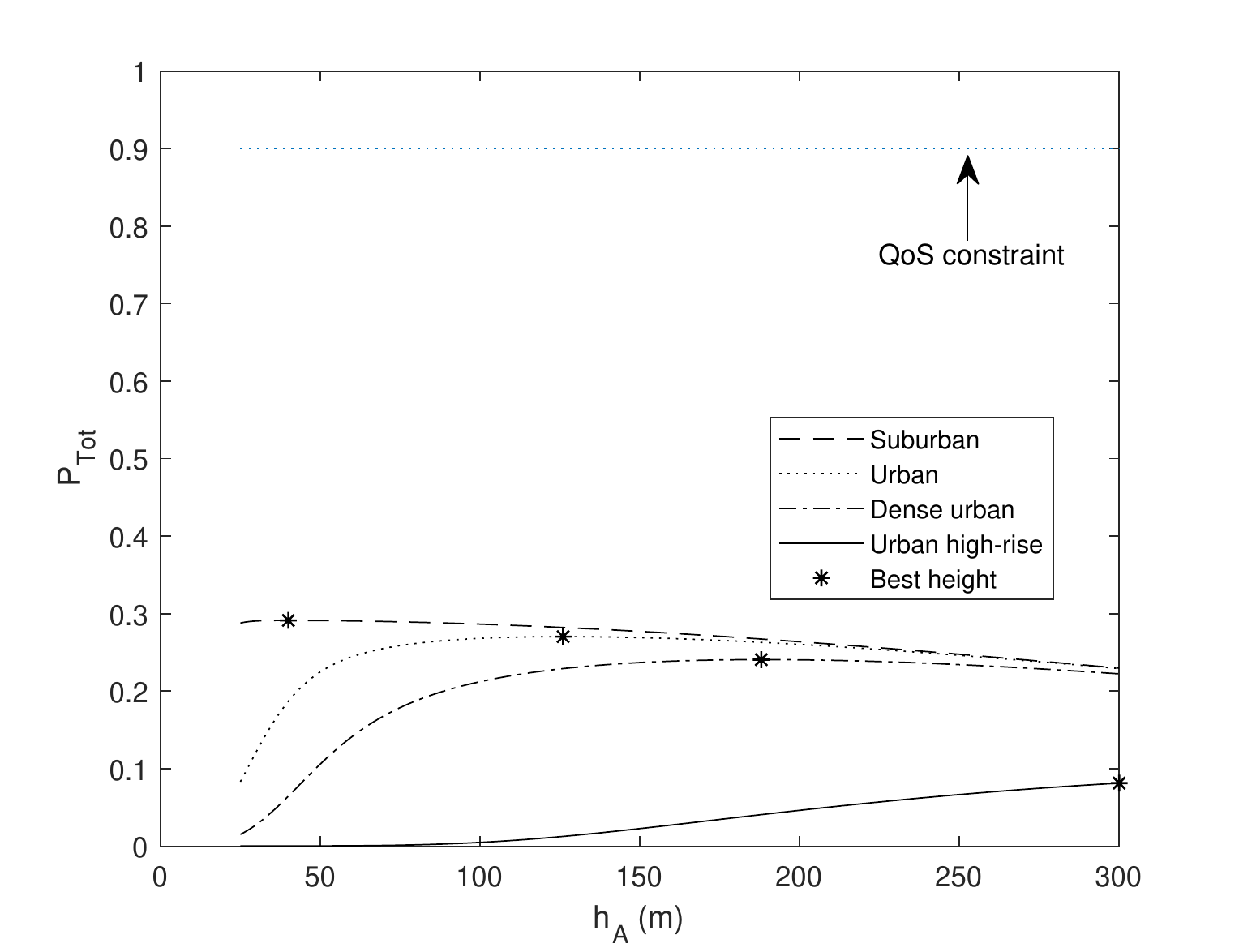}}
\caption{Total rate coverage probability vs. AUE height $\hAUE$ for different environments for (a) $\SINRthresholdAUE=20$ dB and $n=1$, (b) $\SINRthresholdAUE=40$ dB and $n=1$, (c) $\SINRthresholdAUE=20$ dB and $n=10$ and (d) $\SINRthresholdAUE=40$ dB and $n=10$. The best AUE height where $P_{\mathrm{Tot}}$ is maximum is indicated by markers. Dotted line correspond to a $P_{\mathrm{Tot}}$ threshold of $0.9$.}\label{fig:opt}
\end{figure}

Fig. \ref{fig:env} plots $P_{\mathrm{Tot}}$ versus the minimum height to meet QoS constraint for different environments. From Fig. \ref{fig:suburban} for the suburban environment, we can see that due to the lower building density and shorter building height, a QoS of $90\%$ can be achieved by the AUE even if it is flying at a very low height of $25$ m. However, for the rest of the environments (see Fig. \ref{fig:urban}, \ref{fig:dense} and \ref{fig:highrise}), the AUE needs to ascend when AUE flying towards the cell edge, in order to meet the QoS requirement. Moreover, this minimum height increases when the environmental parameters become severer. For an instance, for urban high-rise environment (see Fig. \ref{fig:highrise}), where there is a high density of taller buildings,  a target QoS of $90\%$ cannot be achieved for higher $\SINRthresholdAUE$ and $\dAUEhori$ values even if the AUE is flying at a height of $300$ m. Moreover, the target QoS at $\SINRthresholdAUE=40$ dB can only be satisfied at the first trajectory point (closer to the BS) for urban and dense urban built-up environments, and can not be satisfied at all for urban high-rise environment.

Since the first ($n=1$) and last ($n=10$) trajectory points represent the best and worst cases for the spiral trajectory, we also look at the variation of $P_{\mathrm{Tot}}$ with respect to the height of AUE at these trajectory points. We compare the $P_{\mathrm{Tot}}$ for two different TUE SINR threshold values at $n=1$ and $n=10$ (see Fig. \ref{fig:opt}). At $\SINRthresholdAUE=20$ dB (see Fig. \ref{fig:optH_20n1} and \ref{fig:optH_20n10}), $P_{\mathrm{Tot}}$ increases as the AUE height increases at both trajectory points. However, at $\SINRthresholdAUE=40$ dB (see Fig. \ref{fig:optH_40n1} and \ref{fig:optH_40n10}), there is a best height where the $P_{\mathrm{Tot}}$ is maximized. This trend is prominent when the AUE is closer to the BS (see Fig. \ref{fig:optH_40n1}), and becomes less significant when AUE is closer to the cell boundary (see Fig. \ref{fig:optH_40n10}). Moreover, the best height increases as the environmental conditions become severer for both trajectory points. Note that for $\SINRthresholdAUE=20$ dB, the best height satisfies the QoS constraint of $P_{\mathrm{Tot}}=0.9$, except for urban high-rise at $n=10$. For $\SINRthresholdAUE=40$ dB, the best height satisfies the QoS constraint only for some environments at cell center ($n=1$) and none of the environments at cell edge ($n=10$).

Next, we apply the proposed NOMA scheme to an AUE trajectory inspired by the 3GPP mobility model in \cite{3gpptr36.777}.

\subsection{3GPP Probabilistic LoS and Mobility Models}\label{subsec:3gpp}
\begin{figure}[t]
\centering
\begin{minipage}{.5\textwidth}
  \centering
  \includegraphics[width=1\linewidth]{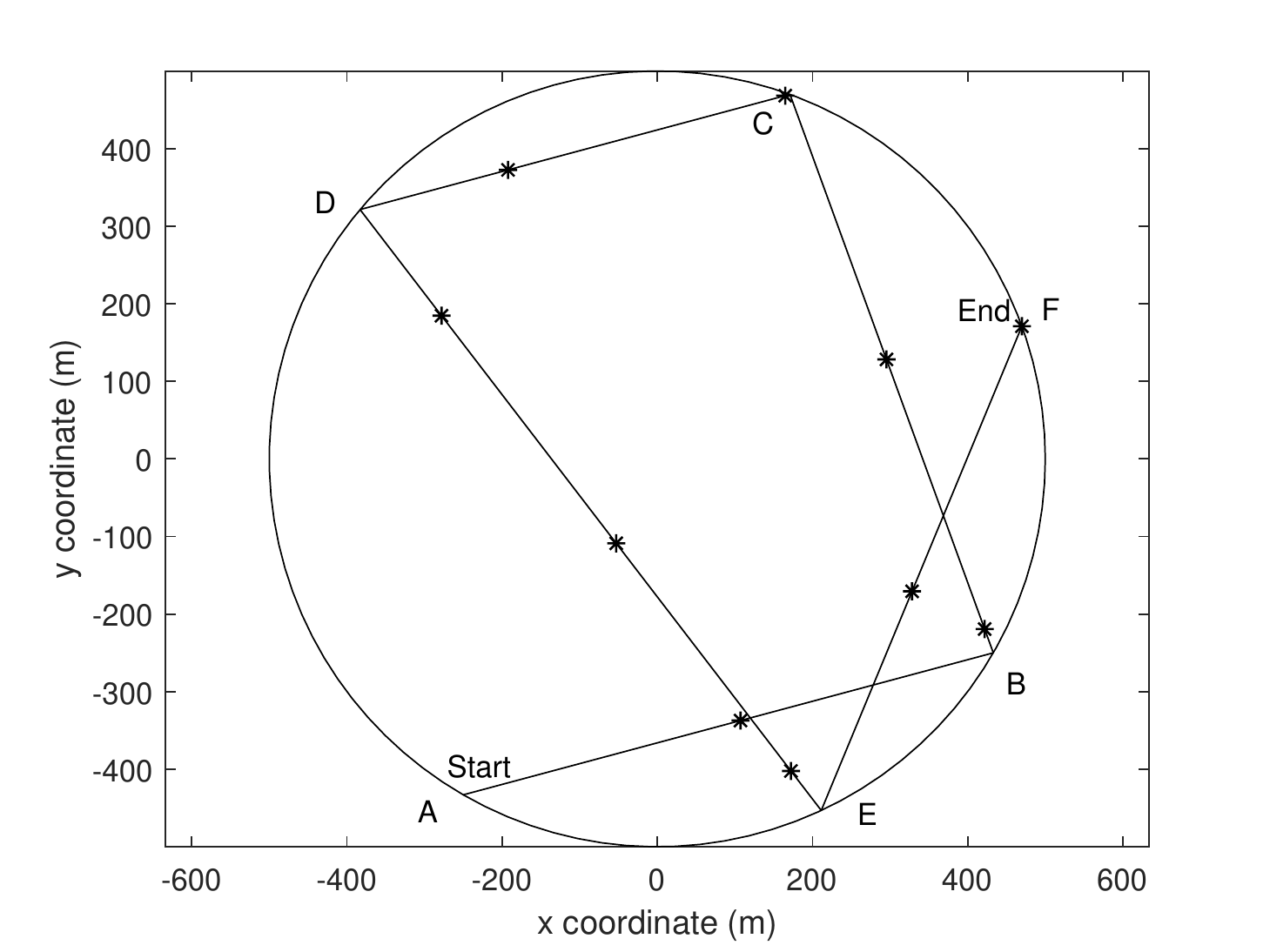}
  \caption{3GPP-inspired trajectory model.}
  \label{fig:3gpp_trajectory}
\end{minipage}%
\begin{minipage}{.5\textwidth}
  \centering
  \includegraphics[width=1\linewidth]{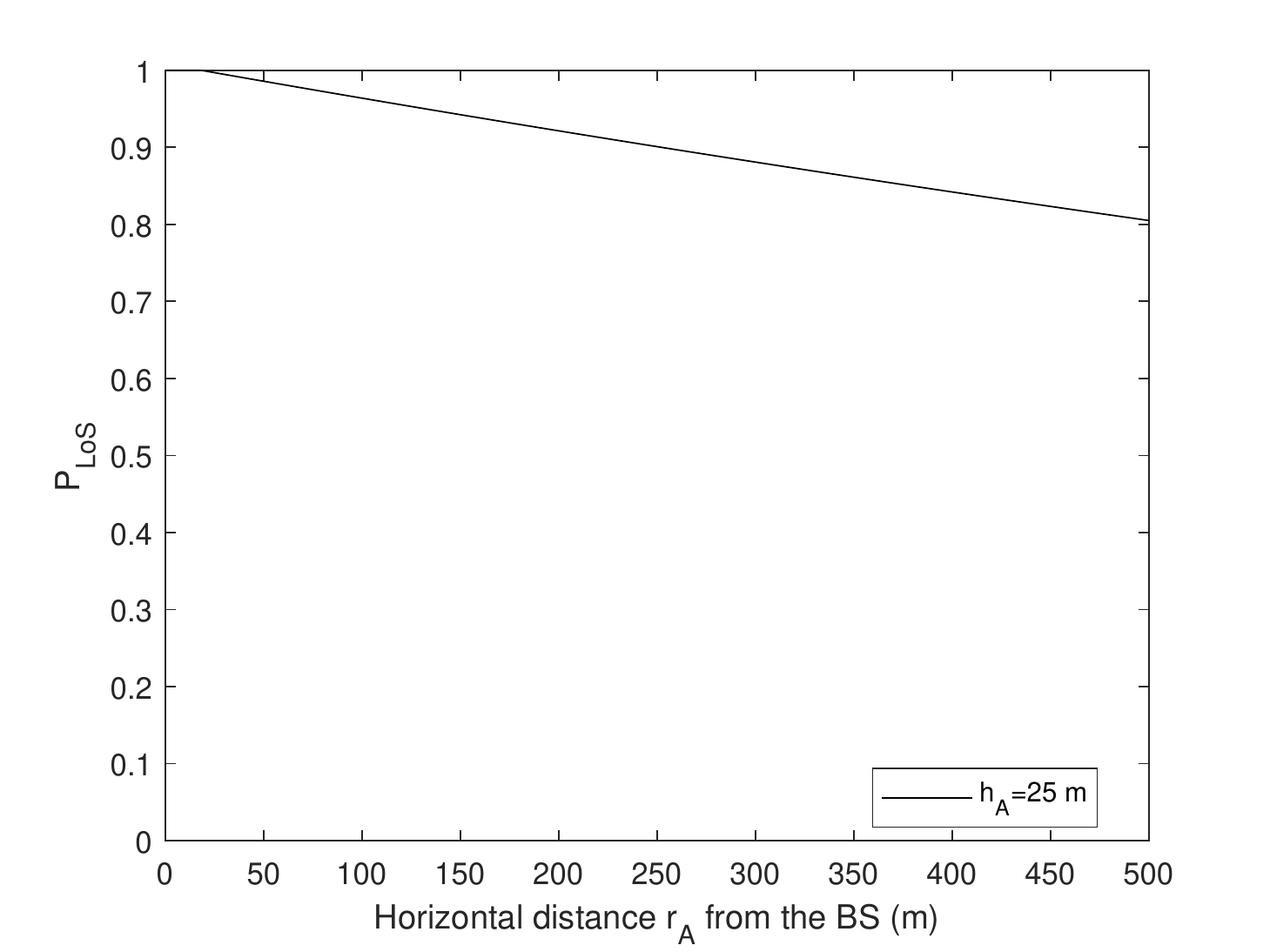}
  \caption{3GPP probabilistic LoS model.}
  \label{fig:3gpp_PLOS}
\end{minipage}
\end{figure}

We adopt the mobility model and probabilistic LoS model proposed in \cite{3gpptr36.777} for this evaluation. In 3GPP probabilistic LoS model, the probability of LoS between the AUE and the terrestrial BS, in the urban environment where the BS antennas are positioned above the rooftop levels of buildings is \cite{3gpptr36.777}
\begin{equation}\label{eq:PLOS_3gpp}
  \mathbb{P}_{LoS}=
    \begin{cases}
      \frac{d_1}{\dAUEhori}+\exp\left(\frac{-\dAUEhori}{p_1}\right)
      \left(1-\frac{d_1}{\dAUEhori}\right), & \mbox{if } \dAUEhori > d_{1}, 22.5 <\hAUE\leq100  \\
      1, & \mbox{if } \dAUEhori\leq d_{1}, 22.5 <\hAUE\leq100 \\
      1, & \mbox{if } 100<\hAUE\leq 300,
    \end{cases}
\end{equation}
\noindent where $d_1=\max(294.05\log_{10}(\hAUE)-432.94,18)$ and $p_1=233.98\log_{10}(\hAUE)-0.95$. The probability of LoS at $\hAUE=25$ m is illustrated in Fig. \ref{fig:3gpp_PLOS}. Compared with ITU LoS probability for the same height from Fig. \ref{fig:PLOS}, we can see that the LoS probability is very favorable even at low height of AUE.

The trajectory model for this scenario is defined as follows. The AUE starts its trajectory at a random initial location at the cell edge. Then, it moves at a randomly selected orientation in a straight line until it reaches the cell edge. Once it reaches the cell border, it picks another random orientation and moves in a straight line, and this process is continued until the AUE reaches its final trajectory point. An example realization, used for the purposed of generating results, is shown in Fig. \ref{fig:3gpp_trajectory}. We consider AUE moves along multiple chords in the cellular cell, where $\mathrm{A},$ $\mathrm{B},$ $\mathrm{C},$ $\mathrm{D},$ $\mathrm{E}$, and $\mathrm{F}$ correspond to the points at the cell edge where AUE changes the orientation of the trajectory. We consider that AUE transmits $10$ times along its trajectory (see Fig. \ref{fig:3gpp_trajectory}).

\begin{figure}[t]
\centering
\begin{minipage}{.45\textwidth}
  \centering
  \includegraphics[width=1\linewidth]{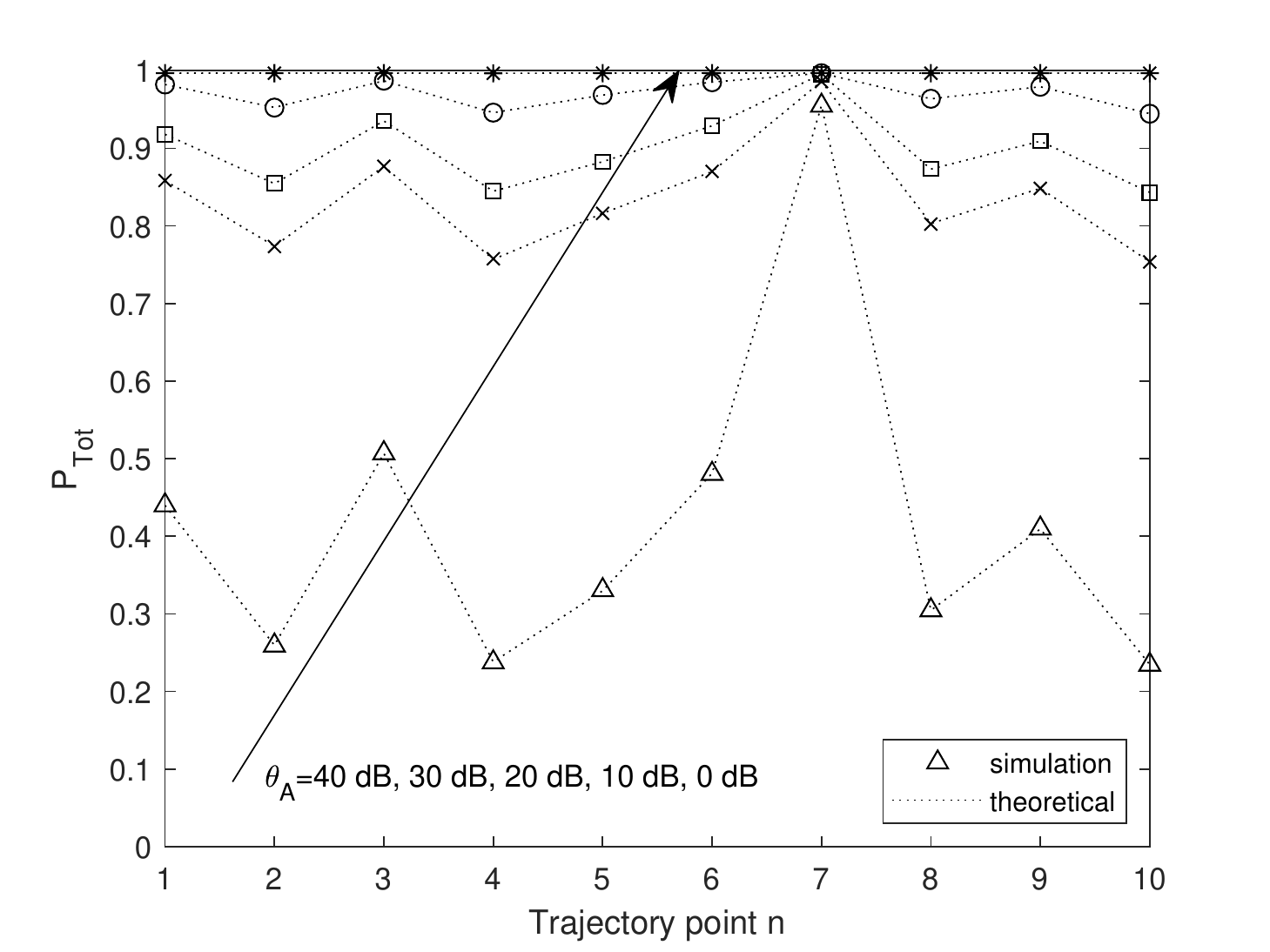}
  \caption{$P_{\mathrm{Tot}}$ for 3GPP-inspired trajectory model and probabilistic LoS model. The simulation values and the theoretical values are represented by bullets and dotted lines, respectively.}
  \label{fig:PT_25_valid_3gpp}
\end{minipage}%
\hspace{1 cm}
\begin{minipage}{.45\textwidth}
  \centering
  \includegraphics[width=\linewidth]{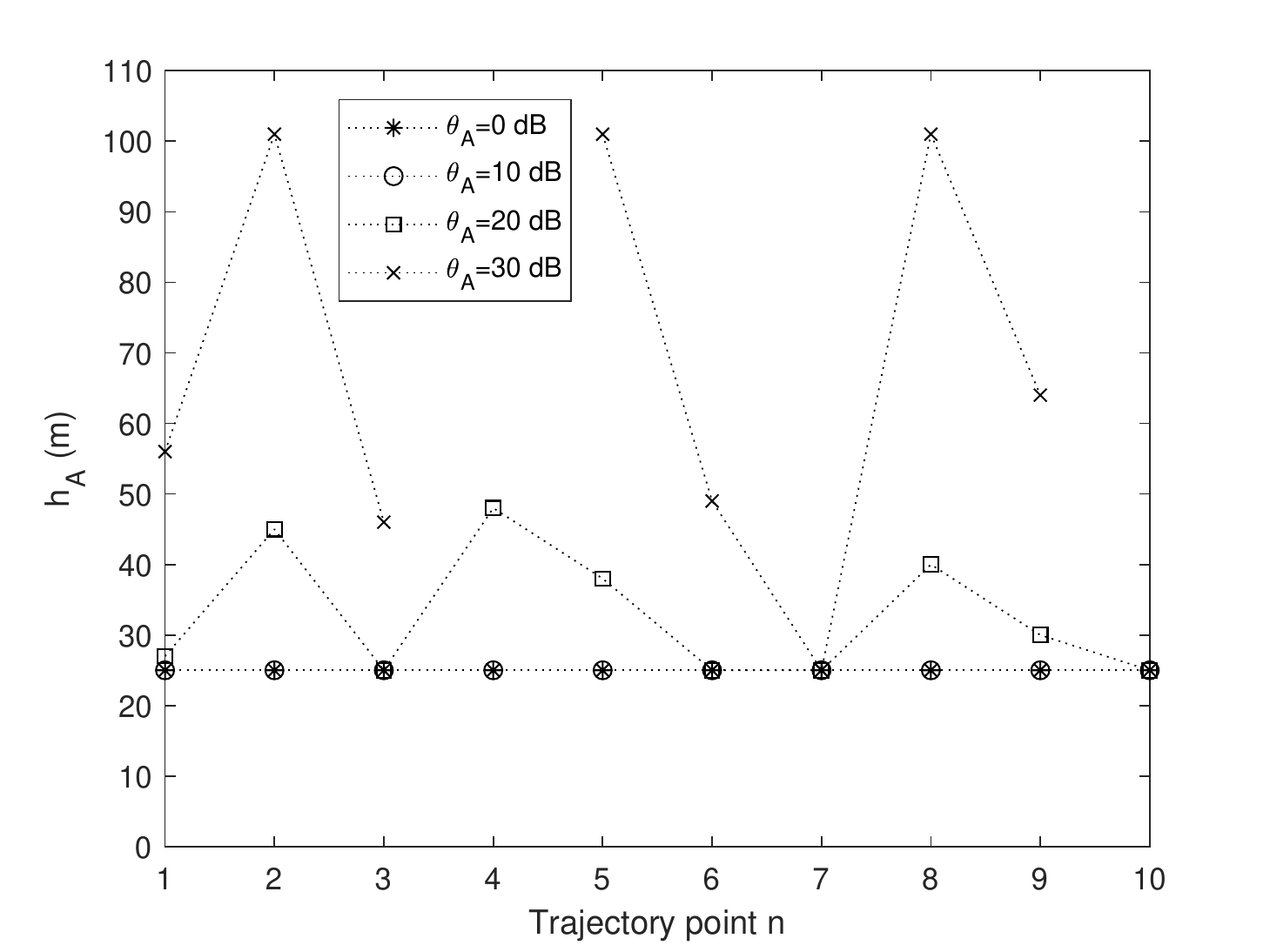}
  \caption{The minimum height of AUE to achieve a target QoS of $90\%$ at each trajectory point in the urban environment for 3GPP-inspired trajectory model and probabilistic LoS model.}
  \label{fig:min_heights_3gpp}
\end{minipage}
\end{figure}

Fig. \ref{fig:PT_25_valid_3gpp} presents the model validation results for the 3GPP probabilistic LoS and mobility model. We can see that simulation results match well with the analytical results. This confirms that our proposed NOMA model can be applied to any trajectory model. We can see that the trajectory point at $n=7$, has the highest total rate coverage probability. This is because at this point the AUE has close proximity to the terrestrial BS, and this is consistent with the trends discussed in the previous trajectory model. Fig. \ref{fig:min_heights_3gpp} illustrates the minimum height of AUE that satisfies a target QoS requirement of $90\%$. Note that, at $\SINRthresholdAUE=30$ dB, trajectory points $n=4$ and $n=10$, fail to meet this QoS requirement within the given height range. This is due to the fact that these transmission points are located very close to the cell edge. We can see that for $\SINRthresholdAUE=20$ and $30$ dB, the minimum height required to maintain a QoS of $90\%$ increases and decreases at various points in the trajectory. This trend is different from the Archimedes' spiral trajectory where the minimum height increases as the AUE moves from the center to the cell edge. This highlights the importance of considering the specific UAV mobility in the modeling and design of UAV communication systems.

\section{Conclusion}\label{sec:conclusion}
In this paper, we considered the coexistence of an AUE with a paired TUEs in a cellular network. We assumed that the AUE flies in a given trajectory path and transmits to the BS periodically.  To facilitate the concurrent uplink transmissions of the AUE and the TUE, we used an aerial-terrestrial NOMA scheme with SIC at the BS. We formulated an analytical framework that evaluates the rate coverage probability of each user and the system, at each transmission point on the trajectory. The numerical results showed that, for the spiral trajectory, the rate coverage probabilities decrease as the target data rate of the AUE increases and the AUE moves away from the BS. We also found the minimum height of AUE at each trajectory point in order to meet a QoS of $90\%$ for different built-up environments. In the spiral trajectory, it was observed that the minimum height increases as the environmental parameters become more severe and when AUE moves towards the cell edge. For the trajectory model adopted from the 3GPP recommendations, it was observed that the minimum height increases and decreases depending on the distance of AUE from the BS.

\begin{appendices}
\section{\underline{Proof of Lemma \ref{lem:pdf_dG}}\label{app:proof_lemma_dG}}
The horizontal distance $\dGUEhori$ between the active TUE and BS at the time of transmission follows an uniform distribution due to the fact that $\xGUE$ and $\yGUE$ are uniformly distributed in the circular cell. Thus, the pdf of $\dGUEhori$ is $f_{\dGUEhori}(r)=\frac{2r}{\cellrad^2}$. The 3D distance $\dGUEeucl$ between the BS and TUE is given by $\dGUEeucl=\sqrt{\hBS^2+\cellrad^2}$. Therefore, the pdf of $\dGUEeucl$ is derived as

\begin{align}
  f_{\dGUEeucl}(z) &=\frac{dr}{dz}f_{\dGUEhori}(r)
   = \frac{d(\sqrt{z^2-\hBS^2})}{dz}f_{\dGUEhori}(\sqrt{z^2-\hBS^2})
   = \frac{z}{\sqrt{z^2-\hBS^2}}\frac{2\sqrt{z^2-\hBS^2}}{\cellrad^2}=
   \frac{2z}{\cellrad^2}.
\end{align}

\section{\underline{Proofs of Lemmas \ref{lem:pdf_SINRG} and \ref{lem:pdf_SINRA}}}\label{app:proof_lemma_xG_xA}

In this appendix, we provide proofs of Lemmas \ref{lem:pdf_SINRG} and \ref{lem:pdf_SINRA}.
\subsection{\underline{Proof of Lemma \ref{lem:pdf_SINRG}}}

The CDF of $\GUERecPow$ can be written as
\ifCLASSOPTIONpeerreview
\begin{subequations}\label{eq:cdf_SINRG_proof}
\begin{align}
  \nonumber
  F_{\GUERecPow}(x)&=\mathbb{P}_{\dGUEeucl,\fadingGUE}\left(\GUETranPow
  \dGUEeucl^{-\alphaGUE}\fadingGUE\GUEantenna<x\right)=
  \mathbb{E}_{\dGUEeucl}\left[\mathbb{P}\left(\fadingGUE<\frac{x}{\GUETranPow
  \dGUEeucl^{-\alphaGUE}\GUEantenna}\right)\right]\\
  \label{eq:cdf_SINRG_proof_fading}
  &=\mathbb{E}_{\dGUEeucl}\left[1-\exp\left(\frac{x}{\GUETranPow
  \dGUEeucl^{-\alphaGUE}\GUEantenna}\right)\right]\\ \label{eq:cdf_SINRG_proof_expect}
  &=1-\int_{\hBS}^{\sqrt{{\hBS}^2+{\cellrad}^2}}\exp\left(\frac{-x}
  {\GUEavgRpow z^{\alphaGUE}z^{-\alphaGUE}\GUEantenna}\right)\left(
  \frac{2z}{{\cellrad}^2}\right)dz=1-\exp\left(\frac{-x}{\GUEavgRpow
  \GUEantenna}\right),
\end{align}
\end{subequations}
\else
\begin{subequations}\label{eq:cdf_SINRG_proof}
\begin{align}
  \nonumber
  F_{\GUERecPow}(x)&=\mathbb{P}_{\dGUEeucl,\fadingGUE}\left(\GUETranPow
  \dGUEeucl^{-\alphaGUE}\fadingGUE\GUEantenna<x\right)\\ \nonumber
  &=\mathbb{E}_{\dGUEeucl}\left[\mathbb{P}\left(\fadingGUE<\frac{x}{\GUETranPow
  \dGUEeucl^{-\alphaGUE}\GUEantenna}\right)\right]\\
  \label{eq:cdf_SINRG_proof_fading}
  &=\mathbb{E}_{\dGUEeucl}\left[1-\exp\left(\frac{x}{\GUETranPow
  \dGUEeucl^{-\alphaGUE}\GUEantenna}\right)\right]\\ \label{eq:cdf_SINRG_proof_expect}
  &=1-\int_{\hBS}^{\sqrt{{\hBS}^2+{\cellrad}^2}}\exp\left(\frac{-x}
  {\GUEavgRpow z^{\alphaGUE}z^{-\alphaGUE}\GUEantenna}\right)\left(
  \frac{2z}{{\cellrad}^2}\right)dz \\
  &=1-\exp\left(\frac{-x}{\GUEavgRpow
  \GUEantenna}\right),
\end{align}
\end{subequations}
\fi

\noindent where \eqref{eq:cdf_SINRG_proof_fading} comes from the fact that $\fadingGUE$ follows an exponential distribution. \eqref{eq:cdf_SINRG_proof_expect} is the expectation with respect to $\dGUEeucl$, where $\GUETranPow=\GUEavgRpow z^{-\alphaGUE}$ and $f_{\dGUEeucl}(z)=\frac{2z}{\cellrad^2}$. Taking the derivative of $F_{\GUERecPow}(x)$ with respect to $x$ we obtain its PDF.

\subsection{\underline{Proof of Lemma \ref{lem:pdf_SINRA}}}
The CDF of $\AUERecPow$ can be expressed as
\begin{subequations}\label{eq:cdf_SINRA_proof}
  \begin{align}
  \nonumber
    F_{\AUERecPow}(x)&=\mathbb{P}_{\fadingAUE,\dAUEeucl}\left(\AUETranPow\pathlossAUE\
    \fadingAUE\AUEantenna<x\right)\\ \nonumber
    &=\mathbb{E}_{\dAUEeucl}\left[\mathbb{P}\left(\fadingAUE<\frac{x}{\AUETranPow
    \pathlossAUE\AUEantenna}\right)\right]\\ \label{eq:cdf_SINRA_proof_fading}
    &=\mathbb{E}_{\dAUEeucl}\Bigg[1-\PLOS\sum_{i=0}^{\mLOS-1}{\left(\frac{\mLOS x}{\AUETranPow\etaLOS{\dAUEeucl}^{-\alphaLOS}\AUEantenna}\right)}^i\frac{1}{i!}\exp\left(
    {\frac{-\mLOS x}{\AUETranPow\etaLOS{\dAUEeucl}^{-\alphaLOS}\AUEantenna}}\right) \\ \nonumber
    &-(1-\PLOS)\sum_{j=0}^{\mNLOS-1}{\left(\frac{\mNLOS x}{\AUETranPow\etaNLOS{\dAUEeucl}^{-\alphaNLOS}\AUEantenna}\right)}^j\frac{1}{j!}\exp\left(
    {\frac{-\mNLOS x}{\AUETranPow\etaNLOS{\dAUEeucl}^{-\alphaNLOS}\AUEantenna}}\right)\Bigg]\\
    &=1-\PLOS\sum_{i=0}^{\mLOS-1}\frac{{\left(\BetaL x\right)}^i}{i!}\exp(-\BetaL x) -
  (1-\PLOS)\sum_{j=0}^{\mNLOS-1}\frac{{\left(\BetaN x\right)}^j}{j!}\exp(-\BetaN x),
  \end{align}
\end{subequations}
\noindent where \eqref{eq:cdf_SINRA_proof_fading} comes from the fact that $\fadingAUE$ follows a Gamma distribution with parameters $\mLOS$ and $\mNLOS$ for LoS and NLoS A2C channel link respectively. Also, note that the expectation in \eqref{eq:cdf_SINRA_proof_fading} is a constant due to the fact that $\dAUEeucl$ is a deterministic variable. $f_{\AUERecPow}(x)$ can be derived by taking the derivative of $F_{\AUERecPow}(x)$ with respect to $x$.

\section{\underline{Proofs of Propositions \ref{prop:P1} and \ref{prop:P2} }}\label{app:proof_prop_p1_p2}
In this appendix, we provide the proofs of Propositions \ref{prop:P1} and \ref{prop:P2}.

\subsection{\underline{Proof of Proposition \ref{prop:P1}}}
The rate coverage probability $P_1$ can be expressed as
\begin{subequations}\label{eq:P1_proof}
  \begin{align}
    \nonumber
    P_1&=\mathbb{P}_{\AUERecPow,\GUERecPow}\left(\frac{\GUERecPow}{\noise}\geq
    \SINRthresholdGUE,\frac{\AUERecPow}{\GUERecPow+\noise}\geq\SINRthresholdAUE\right)\\
    \label{eq:P1_proof_t_wrt_xA}
    &=\mathbb{E}_{\GUERecPow}\left[\mathbb{P}_{\AUERecPow}\left(t\geq\SINRthresholdGUE
    \noise,\AUERecPow\geq\SINRthresholdAUE(t+\noise)\right)\right]\\
    \label{eq:P1_proof_t_expect}
    &=\int_{\SINRthresholdGUE\noise}^{\infty} \mathbb{P}_{\AUERecPow}\left(\AUERecPow
    \geq\SINRthresholdAUE(t+\noise)\right) f_{\GUERecPow}(t)dt\\ \nonumber
    &=\int_{\SINRthresholdGUE\noise}^{\infty} \mathbb{P}\left(\fadingAUE\geq\frac{\SINRthresholdAUE(t+\noise)}
    {\AUETranPow\pathlossAUE\AUEantenna}\right) f_{\GUERecPow}(t)dt\\ \nonumber
    &=\int_{\SINRthresholdGUE\noise}^{\infty} \Biggl(
    \PLOS\sum_{i=0}^{\mLOS-1}\frac{{\left(\BetaL\SINRthresholdAUE\right)}^i}{i!}
    {\left(t+\noise\right)}^i\exp\left(-\BetaL\SINRthresholdAUE(t+\noise)\right)\\
    \label{eq:P1_proof_fadingAUE}
    &+
    (1-\PLOS)\sum_{j=0}^{\mNLOS-1}\frac{{\left(\BetaN\SINRthresholdAUE\right)}^j}{i!}
    {\left(t+\noise\right)}^j\exp\left(-\BetaN\SINRthresholdAUE(t+\noise)\right)
    \Biggl) f_{\GUERecPow}(t)dt\\ \nonumber
    &=\PLOS\sum_{i=0}^{\mLOS-1}\frac{{\left(\BetaL\SINRthresholdAUE\right)}^i}{i!}
    \frac{1}{\GUEavgRpow\GUEantenna}\exp\left(\frac{\noise}{\GUEavgRpow\GUEantenna}
    \right){\left(\BetaL\SINRthresholdAUE+\frac{1}{\GUEavgRpow\GUEantenna}\right)}^
    {-i-1} \\ \nonumber
    &\times\Gamma\left(i+1,\frac{(1+\SINRthresholdGUE)(1+\BetaL\SINRthresholdAUE
    \GUEavgRpow\GUEantenna)\noise}{\GUEavgRpow\GUEantenna}\right)+
    (1-\PLOS)\sum_{j=0}^{\mNLOS-1}\frac{{\left(\BetaN\SINRthresholdAUE\right)}^j}{j!}
    \frac{1}{\GUEavgRpow\GUEantenna}\exp\left(\frac{\noise}{\GUEavgRpow\GUEantenna}
    \right)\\ \label{eq:P1_proof_integrate}
    &\times{\left(\BetaN\SINRthresholdAUE+\frac{1}{\GUEavgRpow\GUEantenna}\right)}^
    {-j-1}\Gamma\left(j+1,\frac{(1+\SINRthresholdGUE)(1+\BetaN\SINRthresholdAUE
    \GUEavgRpow\GUEantenna)\noise}{\GUEavgRpow\GUEantenna}\right).
  \end{align}
\end{subequations}
\noindent In \eqref{eq:P1_proof_t_wrt_xA} we consider that $t$ denotes a random variable with distribution $f_{\GUERecPow}(t)$ and it is a constant with respect to the random variable $\AUERecPow$. Expectation with respect to $t$, and lower and upper boundaries of $t$ are applied in \eqref{eq:P1_proof_t_expect}. \eqref{eq:P1_proof_fadingAUE} comes from the fact that the fading of AUE has a Gamma distribution and, \eqref{eq:P1_proof_integrate} is obtained by integrating \eqref{eq:P1_proof_fadingAUE} with respect to $t$. Finally, \eqref{eq:P1} can be derived by substituting $\muG=\GUEavgRpow\GUEantenna$ into \eqref{eq:P1_proof_integrate}.

\subsection{\underline{Proof of Proposition \ref{prop:P2}}}
The rate coverage probability $P_2$ can be expressed as
\begin{subequations}\label{eq:P2_proof}
  \begin{align}
    \nonumber
    P_2&=\mathbb{P}_{\AUERecPow,\GUERecPow}\left(\frac{\GUERecPow}{\noise}<
    \SINRthresholdGUE,\frac{\AUERecPow}{\GUERecPow+\noise}\geq\SINRthresholdAUE\right) \\ \nonumber
    &=\mathbb{E}_{\GUERecPow}\left[\mathbb{P}_{\AUERecPow}\left(
    t<\SINRthresholdGUE\noise,\AUERecPow\geq\SINRthresholdAUE(t+\noise)\right)\right]\\
    \nonumber
    &=\int_{0}^{\SINRthresholdGUE\noise} \mathbb{P}_{\AUERecPow}\left(
    \AUERecPow\geq\SINRthresholdAUE(t+\noise)\right) f_{\GUERecPow}(t)dt\\ \label{eq:P2_proof_xA_prob}
    &=\int_{0}^{\SINRthresholdGUE\noise} \Biggl(
    \PLOS\sum_{i=0}^{\mLOS-1}\frac{{\left(\BetaL\SINRthresholdAUE\right)}^i}{i!}
    {\left(t+\noise\right)}^i\exp\left(-\BetaL\SINRthresholdAUE(t+\noise)\right)\\
    \label{eq:P2_proof_integrate}
    &+
    (1-\PLOS)\sum_{j=0}^{\mNLOS-1}\frac{{\left(\BetaN\SINRthresholdAUE\right)}^j}{i!}
    {\left(t+\noise\right)}^j\exp\left(-\BetaN\SINRthresholdAUE(t+\noise)\right)
    \Biggl) f_{\GUERecPow}(t)dt,
  \end{align}
\end{subequations}
\noindent where the derivation of $P_{\AUERecPow}(\SINRthresholdAUE(t+\noise))$ in \eqref{eq:P2_proof_xA_prob} is same as that in \eqref{eq:P1_proof}. \eqref{eq:P2} can be obtained by integrating \eqref{eq:P2_proof_integrate} with respect to $t$.

\end{appendices}


\end{document}